\documentstyle[11pt]{article}

 \newcommand{\Bbb}{\bf}
{\catcode `\@=11 \global\let\AddToReset=\@addtoreset}
\AddToReset{equation}{section} 

\newcommand{\thetag}[1]{(#1)}

\newcommand{\al}{\alpha}
\newcommand{\ga}{\gamma}
\newcommand{\de}{\delta}
\newcommand{\cD}{{\cal D}}

\newcommand{\th}{\theta}

\newcommand{\rh}{\rho}
\newcommand{\si}{\sigma}
\newcommand{\ta}{\tau}

\newcommand{\ps}{\psi}

\newcommand{\Si}{\Sigma}

\newcommand{\eq}[1]{(\ref{#1})}

\newcommand{\myspace}{\,}

\newcommand{\commentout}[1]{}  
 
 \newcommand{\be}{\begin{equation}}

\newcommand{\etc}{{\em etc}}

\newcommand{\cV}{{\cal V}}
\newcommand{\cU}{{\cal U}}
\newcommand{\cW}{{\cal W}}
\newcommand{\cH}{{\cal H}}
\newcommand{\ds}{{\cal D}(\Sigma)}
\newcommand{\pds}{\partial\ds}

\newcommand{\ods}{\stackrel{\circ}{\cal D}\!\!(\Sigma)}
\newcommand{\oB}{\stackrel{\circ}{\!\cal B}\!\!}

\newcommand{\ee}{\end{equation}} \newcommand{\bea}{\begin{eqnarray}}
\newcommand{\eea}{\end{eqnarray}}
\newcommand{\beaa}{\begin{eqnarray*}}
\newcommand{\eeaa}{\end{eqnarray*}}

\newcommand{\D}{{\cal D}} \newcommand{\V}{{\cal V}}
\newcommand{\R}{{\Bbb R}} \newcommand{\C}{{\Bbb C}}
\newcommand{\Z}{{\Bbb Z}} \newcommand{\N}{{\Bbb N}}
\newcommand{\RE}{{\Re}} \newcommand{\IM}{{\Im}}
\newcommand{\proof}{\noindent{\bf Proof:}\ }
\newcommand{\Proof}{\noindent{\bf Proof:}\ }

\newcommand{\cf}{{\em cf.\/}\ }
\newcommand{\eg}{{\em e.g.}}
\newcommand{\cfeg}{{\em cf., e.g.,}\ }
\newcommand{\group}{\phi}

\newcommand{\Lie}{\mbox{{{\bf g}}}} 

     \newtheorem{Theorem}   {Theorem}   [section]
     
     \newtheorem{Lemma}     [Theorem]   {Lemma}
     \newtheorem{Proposition} [Theorem] {Proposition}

\begin{document}

\title{Strong cosmic censorship in vacuum space--times with compact, locally
  homogeneous Cauchy surfaces} \author{Piotr T.\
  Chru\'sciel\thanks{Alexander von Humboldt fellow.  On leave
    of absence from the Institute of Mathematics, Polish Academy of
    Sciences, Warsaw.  Supported in part by a Polish Research Council
    grant KBN 2 P302 095 06.
Present address: D\'epartement de Math\'ematiques, Facult\'e des
Sciences, Parc de Grandmont, F 37200 Tours, France.
 {\em e--mail}: chrusciel@Univ--Tours.fr
    }~{}\thanks{Supported in part by the Federal Ministry of Science and
Research, Austria. }   \\ Alan D.\ Rendall\thanks{
Present address: Institut des Hautes Etudes Scientifiques, 35 Route
de Chartres, 91440 Bures sur Yvette, France.
  {\em e--mail}: rendall@ihes.fr
  }~{}$^{\dagger}$ \\ \\Max Planck Institut f\"ur
  Astrophysik\\ Karl Schwarzschild Strasse 1\\ D 85740 Garching bei
  M\"unchen }

\date{gr-qc/9410040}

\maketitle

\begin{abstract}
  We consider the question of strong cosmic censorship in spatially
  compact, spatially locally homogeneous vacuum models. We show in particular
  that strong cosmic censorship holds in Bianchi IX vacuum
  space--times with spherical spatial topology. \end{abstract}

\section{Introduction}
It is widely believed that an important question in classical
general
relativity is that of {\em strong cosmic censorship} (SCC), due to Penrose
\cite{PenroseSCC}.  A mathematical formulation thereof, essentially due
to Moncrief and Eardley \cite{EM} ({\em cf.\/} also \cite{SCC,ChrCM,JimCM}),
is the following:
\begin{quote}
  {\em Consider the collection of initial data for, say, vacuum
   space--times, with the initial data surface
    $\Sigma$ being compact, or with the initial data
    $(\Sigma,\gamma,K)$ --- asymptotically flat. For generic such
    data the maximal globally hyperbolic development thereof is
    inextendible\footnote{When talking about extensions, unless
explicitly specified otherwise we do not assume that the extension
satisfies any field equations.}.}
\end{quote}
The failure of the above would mean a serious lack of predictability of
Einstein's equations, an unacceptable feature of a physical theory.

Because of the difficulty of the strong cosmic censorship problem, a
full understanding of the issues which arise in this context seems to
be completely out of reach at this stage. For this reason there is
some interest in trying to understand that question under various
restrictive hypotheses, {\em e.g.}, under symmetry hypotheses. Such a
program has been undertaken in \cite{EM}, and some further results in
the spatially compact case have been obtained in
\cite{ChIM,IMGowdy,SCC,BChM}. Interestingly enough, the SCC issue
remained open in the class of vacuum\footnote{{\em cf.\/}
  \cite{CollinsEllis} for some results for Einstein --
  perfect fluid  space--times and \cite{RendallBianchiMatter} for
  results on non-vacuum spacetimes with more general matter fields.}
  spatially compact,
  spatially homogeneous
  space--times because of the difficulties in understanding the global
  dynamical behaviour of the Bianchi IX models. The aim of this paper
  is to fill this gap. The first main result of our paper is the
  following:

\begin{Theorem} \label{T1}
  Strong cosmic censorship holds in the class of vacuum Bianchi IX
  space--times with $L(p,1)$, $p=1,2$ spatial topology.
\end{Theorem}

It follows immediately from point 2 of Proposition \ref{Ptop},
Section \ref{BIX} below, that the result is wrong for spatial topology
$L(p,1)$, $p>2$.

Recall that the standard way of proving SCC ({\em cf.\ e.g.\/}
\cite{ChIM}) is to prove uniform curvature blow--up of the metric for
all but a non--generic set of initial data. Now one expects this to be
true for Bianchi IX models because of the apparently ergodic behaviour
of the dynamics there, together with the uniform volume contraction as
the ``boundary'' of the maximal globally hyperbolic development
is approached
({\em cf.\ e.g.\/} \cite{Hobill} and references therein for various
results supporting this point of view; {\em cf.\ e.g.\/}
\cite{VinceGrubisic,Cantopoulos1,Cantopoulos2,CRughsquare} for some results
suggesting that there might still be some surprises left on the road
to the understanding of the dynamical behaviour of Bianchi IX models).
Because of the complicated dynamical behaviour of the Bianchi IX
models no rigorous proofs of curvature blow up have been published,
and in fact it seems that no one has ever considered this question
from a numerical point of view. Here we avoid this question by
isolating directly those initial data which lead to the formation of a
Cauchy horizon, leaving aside the problem of the long--time behaviour
of the metric for those data which do not form a Cauchy horizon. This
latter question has an interest of its own, and it certainly deserves
further investigation. Fortunately, it is not needed in our proof of
SCC for Bianchi IX models.

It should be pointed out that a version of Theorem \ref{T1} has been
proved by Siklos in \cite{SiklosCMP}. In that reference analyticity of
the space--time is assumed. Moreover it is assumed there that the
action of the isometry group on the globally hyperbolic part of the
space--time analytically extends to the Cauchy horizon. The main work in
this paper is to show that these hypotheses can be removed. We wish to
acknowledge  that in several places we borrow heavily on the
arguments presented in \cite{SiklosCMP}. Some of the arguments here
are perhaps somewhat simpler than the corresponding ones in
\cite{SiklosCMP}; moreover we have found it necessary to reorganize
Siklos' arguments in various places, because of different assumptions
made. For those reasons we have found it useful to give a detailed
exposition of the whole proof, which overall is rather similar to that
of \cite{SiklosCMP}.

Recall that the issue of strong cosmic censorship arises because of
possible non--uniqueness of solutions of the Einstein  equations beyond
Cauchy horizons.  This is because any Cauchy data $(\Sigma, \gamma,
K)$ define a unique (up to isometry) maximal globally hyperbolic
development $(M, g)$, but whenever $(M, g)$ is extendible
uniqueness
of the vacuum extensions is lost, at least if one does not impose some
further restrictions.  In fact, in Section \ref{BIXV} we present
a construction which shows that this occurs in any space--time with a
Cauchy horizon, even when analyticity conditions are imposed on the
extended space--time. The
Taub--NUT space--times are a well known example with the property that
every Taub--NUT initial data admit at least two maximal (vacuum) developments
(\cf \cite{ChImaxTaubNUT}).  These
developments will be called the standard Taub--NUT space--times.
Now the essential difference between those two space--times is in
the
way the boundaries $\partial{\cal D}(\Sigma)$ are
``glued" to the globally hyperbolic Taub region, and it is natural to
ask how many ways of doing this glueing exist. Our next result here
shows that the standard Taub--NUT space--times exhaust all the
possibilities:

\begin{Theorem} \label{T4} Let $(M,g)$ be a vacuum
  Bianchi IX space--time with a partial Cauchy surface $\Sigma $ such
  that $\D(\Sigma ;M)$ is maximal globally hyperbolic, with both the
  future Cauchy horizon $\cH^+(\Sigma ; M)$ and the past Cauchy
  horizon $\cH^-(\Sigma ; M)$ --- connected, and $\cH^+(\Sigma ;
  M)\cup \cH^-(\Sigma ; M)$ --- {\em     not} empty. 
  Then there
  exists a standard Taub--NUT space--time $(\hat{M}, \hat g)$ and an isometric
  diffeomorphism $i$
  $$ i :\,\,\stackrel{\circ}{\cal D} (\Sigma ; M) \rightarrow \,\,\,
  \stackrel{\circ}{\cal D} (i(\Sigma ); \hat{M})\,, $$
  such that $i$ extends by continuity to a one--to--one map $\bar{\imath}$ $$
  \bar{\imath} : \overline{\D(\Sigma ; M)} \rightarrow
  \overline{\D(i(\Sigma );\hat{M})}.
$$ Moreover $\bar \imath$ is a diffeomorphism of manifolds with boundary
between $\overline{\D(\Sigma ; M)}$ and $\bar \imath(\overline{\D(\Sigma;
  M)})$.
\end{Theorem}
Here $\,\,\stackrel{\circ}{\cal D}(\Sigma ; M) $ denotes the interior of the
domain
of dependence ${\cal D}(\Sigma ; M) $ of a partial Cauchy surface
$\Sigma$, while $\bar\Omega$ denotes the closure of a set $\Omega$.

In addition, in Section \ref{BIXV} we prove that the only maximal
Taub--NUT space--times on which $G=SO(3)$ or $G=SU(2)$ acts by
isometries are the standard Taub--NUT space--times, {\em cf.\/}
Theorem \ref{TBV.I}.  It would be of interest to find some other
conditions, weaker than the above, which single out the standard
Taub--NUT space--times in the collection of all the extensions of the
globally hyperbolic region of the Taub--NUT space--times.

Let us remark that in Theorem \ref{T4} the condition that each of
$\cH^+(\Sigma ; M)$ and $\cH^-(\Sigma ; M)$ are connected is
necessary. Indeed, counterexamples with non--connected, say,
$\cH^+(\Sigma ; M)$, can be constructed by making a ``left future
extension" for some set of null geodesics in Taub space--time, and a
``right future extension" for another set of those. (The Cauchy
horizon will of course not be compact here -- clearly a compact, say,
future, Cauchy horizon in a Bianchi IX space--time has to be
connected.)

An interesting class of space--times in which to study the question
of
strong cosmic censorship is that of space--times evolving out of {\em
  locally homogeneous} initial data.
Recall that a homogeneous geometry $(\Sigma,\gamma)$ is defined by the
existence of a group $G$ which acts transitively on $\Sigma$ by
isometries of $\gamma$.  Loosely speaking, local homogeneity is then
defined by the requirement of the existence of a locally transitive
action by isometries, {\em cf.\/} the beginning of Section \ref{LH}
for a precise definition.  Recall that in the general
relativists' terminology every space--time with locally homogeneous
spacelike surfaces can be assigned a {\em symmetry type}, namely one
of the 9 Bianchi types, and the
Kantowski--Sachs\footnote{\label{KSfootnote}Following the standard
  terminology, the Kantowski--Sachs symmetry type is defined here by
the requirement of local homogeneity together with the
  condition that there is no three dimensional isometry group acting
transitively
  on the universal covering space. This implies that the universal
  cover is $\R\times S^2$, and the connected component of the identity
  of the isometry group of the universal cover is $\R\times SO(3)$
  with the obvious action.} type, {\em cf.\ e.g.\/}
\cite{KoikeTanimotoHosoya,FIK,Scott,Thurston,IsenbergJackson,Malcolm}.
It turns out that the methods needed to analyze the Bianchi IX case
carry over without any essential modifications to the case of locally
homogeneous initial data.  Let us first mention the following result
which  shows   that there could potentially be a SCC problem in the class of
  space--times with locally homogeneous Cauchy surfaces:

\begin{Proposition}\label{P0}
\begin{enumerate}
  \item For every symmetry type except\footnote{We use the conventions of
\cite{SiklosCMP} for labelling the Bianchi groups.}  for Bianchi VI$_0$
 there exists a spatially homogeneous
    vacuum space--time with a non--empty homogeneous Cauchy horizon.

\item Let the symmetry type be one of the following: Kantowski--Sachs,
Bianchi I, II, III=VI$_{-1}$, VII$_0$,   VIII, IX. For every such symmetry type
there exists
a spatially locally homogeneous, {\em spatially compact} vacuum
space--time with a non--empty locally homogeneous Cauchy horizon. No
such space--times exist if the symmetry type is Bianchi IV, V, VI$_h$,
$h\ne -1$, VII$_h$, $h\ne 0$.
\end{enumerate}
\end{Proposition}

Note that the list of Bianchi geometries in the existence part of
point 2 above coincides with the Bianchi A geometries, except for the
type VI$_0$
which
is missing there. The non--existence
part of the list for the spatially compact case consists of Bianchi B
geometries, except for the type III=VI$_{-1}$ for which a spatially
compact model exists.

As discussed at the beginning of Section \ref{LH}, strong cosmic
censorship ``half--fails" in all locally homogeneous Kantowski--Sachs
models. Indeed, those space--times always have a Cauchy horizon, say,
to the future, and a curvature singularity to the past.  Recall now
that SCC is satisfied in {\em homogeneous} Bianchi I models ({\em
  i.e.,} in Bianchi I models with globally defined Killing vectors;
{\em cf.\ e.g.\/} \cite{SCC} for a detailed discussion). Interestingly
enough, it turns out that this is not the case any more when spatially
locally homogeneous models with Bianchi I symmetry type are
considered. In this class of space--times the SCC question turns out
to depend upon the topology of the partial Cauchy surface. This is due
to the fact, that some topologies allow only those locally homogeneous
initial data for which the resulting maximal globally hyperbolic
space--time is extendible. This is discussed in detail in Section
\ref{topology} for the Bianchi I symmetry type.
For the remaining Bianchi models in the discussion of SCC
one needs to analyze case by case all the admissible topologies, as
done in Section \ref{topology} in the Bianchi I case, which is a task
which lies beyond the scope of this paper. (Such an analysis would be
rather lengthy and tedious but probably otherwise straightforward,
basing on the list of topologies given in \cite{KoikeTanimotoHosoya}.)
Nevertheless, as we shall show below, there is a
topology--independent sense in which SCC is satisfied in those models.

When considering the exact known solutions which contain Cauchy
horizons, it is striking that they seem to display more symmetries
than the typical representatives of the families they belong to. This
is {\em e.g.\/} true for Taub--NUT space--times, considered as members
of the Bianchi IX family --- for the former the isometry group is
larger by a factor $U(1)$. This is true in some approximate
sense in the class of polarized Gowdy space--times, in which one of
the functions characterizing the solutions tends to a constant on each
connected component of the Cauchy horizon \cite{ChIM}. Let us also
mention the Isenberg--Moncrief conjecture
\cite{VinceJimcompactCauchy}, that space--times with a compact Cauchy
horizon must have at least one Killing vector. As shown in Section
\ref{LH}, in a
space--time with a locally homogeneous Cauchy surface for every point
there exists a neighbourhood thereof on which a Lie algebra $\Lie$ of
Killing vectors of dimension at least three is defined; $\Lie$ will be
called the {\em local Killing algebra}.  [It turns out that $\Lie$ can
be chosen so that it does not depend upon the choice of points.]  In
our context it is tempting to conjecture that $\Lie$ must be at least
four dimensional whenever a Cauchy horizon occurs.
This turns out to be true in the spatially compact case, and
we have the following rather elegant formulation of SCC in this class
of space--times:

\begin{Theorem} \label{T3}
  Let $(M, g)$ be a vacuum space--time with a compact locally
  homogeneous partial Cauchy surface $\Sigma $ such that $\D(\Sigma)$
  is maximal globally hyperbolic and such that $\partial\D(\Sigma )
  \not= \emptyset$.
Then the local
  Killing vector algebra is at least four--dimensional.
\end{Theorem}

According to Siklos \cite{SiklosCMP}, there exist Bianchi VI$_{-1/9}$
space--times with a Cauchy horizon for which $\Lie$ is only three
dimensional\footnote{ S.  Siklos, private communication.}.  This shows
the necessity of the compactness condition above.

Let us emphasize that from what has been said in \cite{SiklosCMP} it
follows that a generic space--time (in the sense of parameter counting
for simply connected models) of a given Bianchi type will {\em not}
have four Killing vectors, so that Theorem \ref{T3} establishes indeed
some kind of non--genericity of Cauchy horizons in this class of metrics.

This paper is organized as follows. In Section \ref{BIX} we prove
strong cosmic censorship in Bianchi IX space--times.  In Section
\ref{topology} we discuss the question of strong cosmic censorship in
spatially
locally homogeneous models with Bianchi I symmetry type.  In Section
\ref{LH} we consider general space--times with locally homogeneous,
compact partial Cauchy surfaces, and we prove Proposition \ref{P0}
 and Theorem \ref{T3} there. In Section
\ref{BIXV} we consider the question of uniqueness of extensions of the
globally hyperbolic region of the Taub--NUT space--times.

{\bf Acknowledgements} We acknowledge the hospitality of the E.
Schr\"odinger Insitute in Vienna during part of the work on this
paper. P.T.C. also wishes to thank J. Ehlers
and the Max Planck Institute for Astrophysics in Garching for
hospitality. We are grateful to B.  Edgar, G. Galloway, R. Geroch,
G. Hall, J.  Isenberg, S. Siklos, P. Tod and R. Wald for useful
correspondence, or comments, or discussions.

\section{Bianchi IX space--times with Cauchy horizons}
\label{BIX}

In this section we shall prove Theorem \ref{T1}.
Let us start by fixing the
terminology.  $(M,g)$ will be called a vacuum space--time if $M$ is a
four--dimensional manifold\footnote{All manifolds considered in this
  paper are assumed to be smooth, Hausdorff, paracompact, connected
  and orientable.}, and $g$ is a Lorentzian metric on $M$ satisfying
the vacuum Einstein equations (perhaps in a distributional sense).  We
shall moreover assume that $(M,g)$ is time oriented. Here, and
everywhere else, we use the terminology of \cite{HE} unless indicated
otherwise.  Throughout this paper the symbol $\Sigma$ will denote a
partial Cauchy surface in $M$, with $\gamma$ the metric induced from
$g$ on $\Sigma$, and $K$ the extrinsic curvature tensor of $\Sigma$.
$(\gamma, K)$ will always be assumed to satisfy the vacuum constraint
equations.

Recall that given initial data $(\Sigma , \gamma, K)$ for the Einstein
equations there exists ({\em cf. e.g.\/} \cite{CBY}) a maximal
globally hyperbolic development $(M, g)$ of $(\Sigma, \gamma, K)$.
$(M,g)$ is uniquely defined by $(\Sigma, \gamma, K)$ up to isometry.
{\em We shall only consider here space--times which are maximal globally
hyperbolic developments or extensions thereof.} In the case of
a
space--time which is an extension of a maximal globally hyperbolic one
with Cauchy surface $\Sigma$, we shall {\em always\/} assume that
$\Sigma$ remains a partial Cauchy surface in the extended space--time.
In particular $\Sigma$ will {\em always\/} be an achronal spacelike
hypersurface. In our context this condition is equivalent to the
condition of {\em causal regularity} of the extensions used in
\cite{ChImaxTaubNUT}.

Let $\Sigma$ be a compact
three-dimensional manifold and let $(\Sigma, \gamma, K)$ be initial
data for the vacuum Einstein equations which are invariant under an
effective action of $G= SU(2)$ or $G = SO(3)$, with three-dimensional
principal orbits.  A couple $(M, g)$ will be called a Bianchi IX
vacuum space--time if $(M,g)$ is the maximal globally hyperbolic vacuum
development of ($\Sigma, \gamma, K$) or an extension thereof.  (It
follows from, \eg, \cite{FischerinCarmeli} that $\Sigma$ must be a
lens space $L(p, 1)$, $p \in\N$; recall that $L(1, 1) = S^3$ ---
the three-dimensional sphere.)  Let $\,\!\stackrel{\circ}{\cal
  D}\!\!(\Sigma)$ denote the interior of the domain of dependence
$\D(\Sigma)$ of $\Sigma$ in $(M,g)$; we shall write $\D(\Sigma; M)$,
\etc. for $\D(\Sigma)$ when confusions are likely to occur.  As has
been shown in \cite{SCC} (\cf also \cite{Chorbits}) there exists a
smooth action of $G$ on $\,\!\stackrel{\circ}{\cal D}\!\!(\Sigma)$ by
isometries:
 $$ G \ni
g
\mapsto \phi_g : \ \stackrel{\circ}{\cal D}\!\! (\Sigma) \rightarrow
\ \stackrel{\circ}{\cal D}\!\! (\Sigma).  $$ We
have the following (recall that $f$ is in $C^{k,1}$ if the $k$--th
derivatives of $f$ are Lipschitz continuous; this will be the case if,
{\em e.g.\/}, $f$ is in $C^{k+1}$):

\begin{Proposition}\label{P1}
 Let $(M,g)$ be a vacuum Bianchi IX space--time
with a metric $g$ of differentiability class $C^{k,1}$, $k\geq 1$.
Suppose moreover that $\partial\D(\Sigma)\neq \emptyset$, where
$\partial\Omega$ denotes the topological boundary of a set $\Omega$.
Then
\begin{enumerate}
\item There exists an open dense subset of $\partial\D(\Sigma)$ which
  is a submanifold of $M$ of $C^{k,1}$ differentiability
  class.
\item There exist open sets ${\cal O} \subset \partial\D(\Sigma)$ and
  $ \cU \subset G$ such that for $g \in \cU$ the maps ${\group}_g$ extend
  by continuity to differentiable maps $$ \hat{\group}_g : {\cal O}
\rightarrow\partial\D(\Sigma) ,$$ with $\{\hat{\group}_g(p), g \in
{\cU}\}$ --- a three--dimensional subset of $M$.
\end{enumerate}
\end{Proposition}

{\bf Remark:} Note that in general one does not expect a Cauchy horizon
to be significantly better behaved than a Lipschitz hypersurface, even
in space--times with a smooth metric. Here the higher
differentiability of the (open and dense subset of the) Cauchy horizon
follows from the existence of the isometry group. It should be noted
that Proposition \ref{PLH.3} below shows that ${\cal O} =
\partial\D(\Sigma)$, which in turn implies that the whole $ \partial
\D(\Sigma)$
is a submanifold of $M$ of $C^{k,1}$ differentiability class.

\proof\ On $\stackrel{\circ}{{\cal D}}\!\! (\Sigma)$ the metric is
smooth and there any Killing vector $X$ satisfies the equations \be
\nabla_\mu\nabla_\nu X_\sigma = {R^\lambda}_{\mu\nu\sigma}X_\lambda.
\label{EQ.1}\ee
As the metric on $M$ is $C^{1,1}$, $R^\lambda_{\mu\nu\sigma}$ is
uniformly bounded on $\overline{\D(\Sigma)}$ in local coordinates, and
from (\ref{EQ.1}) it easily follows that $X$ extends by continuity to
a vector field $\hat{ X}$ of $C^{k,1}$-up-to-boundary
differentiability class on $\overline{\D(\Sigma)}$.  Let \Lie\ be the
Lie algebra of $G$, by an abuse of notation we shall identify an
element $e$ of \Lie\ with the corresponding Killing vector field
$\hat{X} =\hat{ X}(e)$.

Let $p\in \partial\D(\Sigma)$, there exists a neighbourhood ${\cal
  V}_p$ of the origin in $\Lie$ such that for all $g\in { \cal{U}}_p
\equiv \exp ({\cal V}_p)$ we can define $\hat {\group }_g(p)$ in the
standard way, i.e., if $g= \exp(\hat{X})$, then $\hat{\group }_g(p)$
is defined by following the orbit of $\hat{X}$ through $p$ a unit
distance.  Clearly $\hat{\group }_g$ extends continuously ${\group
  }_g$.  Since ${\group }_g$ maps $\stackrel{\circ}{{\cal D}}\!\!
(\Sigma)$ into $\stackrel{\circ}{{\cal D}}\!\! (\Sigma)$, it follows
that for $g\in {\cal{U}}_p$ the (locally defined) map $ \hat{\group
  }_g$ takes a neighbourhood ${\cal{O}}_p \subset
\partial {\cal{D}}(\Sigma)$ of $p$ into $\partial {\cal{D}}(\Sigma)$.
Because the Killing vectors are of differentiability class $C^{k,1}$
up-to-boundary on $\overline{\D(\Sigma)}$, the maps $\hat{\group
}_g(p)$ are of differentiability class $C^{k,1}$ with respect to $p$.

Let us show that there exists $p \in \partial{ \cal{D}}(\Sigma)$ such
that the set $\{\hat{\group }_g(p), g \in {\cal{U}}_p\}$ is
three--dimensional.  Suppose, first, that for all $p \in
\partial\Sigma$ we have $\{\hat{\group }_g(p), g \in {\cal{U}}_p\} =
\{p\}$.  It follows that for every Killing vector $\hat{X}$ we have
$\hat{X}(p) = 0$ for $p \in \partial{ \cal{D}}(\Sigma)$.
{}From
\cite{PenroseDiffTopo} or \cite{HE} it follows that $\partial{
\cal{D}}(\Sigma)$ is a Lipschitz continuous surface. By a theorem of
Rademacher \cite{FedererMeasureTheory} it follows that $\partial{
\cal{D}}(\Sigma)$ has a tangent plane almost everywhere. Now it is not
too difficult to adapt the arguments of the proof of Lemma 2.1.1 of
\cite{SCC} to cover\footnote{In \cite{SCC} it was assumed
that the hypersurface in question is {\em not} null.  This restriction
is, however, not needed for the argument to go through.} the case of
such hypersurfaces,  to conclude that
$\hat{X} \equiv 0 $ on $M$ for all Killing vectors, which is obviously
not the case.

We have thus shown that $\hat{\group }_g(p)$ cannot act trivially on
$\partial\D(\Sigma)$, hence there exist $p \in \partial\D(\Sigma)$
such that $\hat{\group }_g(p) \not= p$ for some $g \in {\cal U}_p$.
Let $p$ be such a point and define
$$
\hat S_p = \{\hat{X} \in {\cal
  V}_p : \ \hat{\group }_{\exp(\hat{X})}(p) = p\} .
$$
Let $S_p$ be
the linear subspace of \Lie\ spanned by $\hat{ S}_p$ and set $$ G_p =
\exp(S_p).  $$ $G_p$ is a Lie subgroup of $G$ which does not coincide
with $G$, and from the structure of $G$ (recall that $G=SU(2)$ or
$SO(3)$) it follows that either $G_p = \{e\}$, where $e$ is the
identity element of $G$, or $G_p \approx U(1)$.  In the former case
the Killing vectors span a three--dimensional subspace of $T_pM$, and
point 2 follows by setting $ {\cal U} ={\cal U}_p$, $ {\cal O} ={\cal
  O}_p$, and passing to subsets thereof if necessary.  Consider then
the case $G_p \approx U(1)$, let $\hat{X}$ be a generator of $G_p$,
consider the space $(\hat{X})^{\perp_k} \subset \Lie$, where
``$\perp_k$" denotes the orthogonal complement with respect to the
Killing metric $k$ on $G$.  From the structure of $G$ it follows that
for $g \in G_p$ the maps $ad_g$ act on $(\hat X)^{\perp_k}$ by
rotations.  This implies that for $0 \not= \hat{Y} \in (\hat
X)^{\perp_k}$, the vectors $\hat{Y}(p)$ at $p$ are non--vanishing and
spacelike.  Let $$ A_p = \{ X \in T_p M : g (X, \hat{Y}) = 0
\;\mbox{for all}\; \hat{Y} \in (\hat X)^{\perp_k}\} .
$$ As the maps $(\hat{\phi}_g)^*_p$, where $\psi^*_p$ denotes the
pull--back map at p derived from the map $\psi$, preserve the metric
tensor at $p$, it follows that $A_p$ is a two--dimensional timelike
plane invariant under $(\hat{\group }_g)^*_p$, $g\in G_p$.
Consequently, either $G_p$ acts trivially on $A_p$, or it acts on it
by boosts.  In the latter case we can find a nearby point $p'$ on the
null generator of $\partial\D(\Sigma)$ through $p$ such that the
Killing vectors span a three--dimensional subspace of $T_{p'} M$, and
setting ${\cal U}= {\cal U}_{p'}$, and perhaps passing to subsets
thereof if necessary point 2 follows.  The former case is impossible,
which can be seen as follows: if $(\hat\phi_g)^*_p$ acts trivially on
$A_p$, let $X_p \in A_p$ be any null vector at $p$ which is not
tangent to the generator of $\partial\D(\Sigma)$, let $\hat\Gamma$ be
a maximally extended null geodesic through $p$ which is tangent to
$X_p$, set $\Gamma = \hat\Gamma \cap \stackrel{\circ}{{\cal D}}\!\!
(\Sigma)$.  If $\Gamma$ were not empty, it would be a null geodesic in
$\stackrel{\circ}{{\cal D}}\!\!  (\Sigma)$ invariant under $G_{p'}$
but there are no null geodesics in $\stackrel{\circ}{{\cal D}}\!\!
(\Sigma)$ invariant under any non--trivial element of $G$.

We have thus shown that there exists $p \in \partial \D(\Sigma)$ such
that the Killing vectors span a three--dimensional plane in $T_pM$.
Indeed, the argument above shows that the set of such $p$'s is open
and dense in $\partial \D(\Sigma)$.
Because $\partial\D(\Sigma)$ is invariant under the (local) action
$\hat{\group }_g$ for $g \in {\cal U}_p = \cal{U}$, it follows that
there exists a neighbourhood ${\cal O} = {\cal O}_p \subset \partial
\D(\Sigma)$ of $p$ such that $\partial\D(\Sigma)$ is a submanifold of
$M$ of differentability class $C^{k,1}$, with the action of $\cal{U}$
on $\cal O$ being simply transitive. \hfill $\Box$

When $\partial{\cal D}(\Sigma)$ is compact, the following simpler
proof of Proposition \ref{P1} can be given. Arguing as above, the
Killing vector fields can be extended by continuity to vector fields on
$\overline{{\cal D}(\Sigma)}$, tangent to $\partial{\cal D}(\Sigma)$.
By compactness of $\partial{\cal D}(\Sigma)$ every vector field there
is complete, so that an action of $G$ is defined on $\partial{\cal
  D}(\Sigma)$. By transitivity of the action of $G$ on Cauchy
surfaces, and by the fact that a sequence of Cauchy surfaces
accumulating at $\partial{\cal D}(\Sigma)$ exists, it follows that the
action is transitive on $\partial{\cal D}(\Sigma)$. From the list of
actions of $G$ given in \cite{FischerinCarmeli} it follows that the action is
free, and by the arguments given above $\partial{\cal D}(\Sigma)$ is a
differentiable submanifold of differentiability class $C^{k,1}$.

The orbits of the group action on $\stackrel{\circ}{{\cal D}}\!\!
(\Sigma)$ form a smooth foliation.  Denote the future--pointing unit
normal to this foliation by $u$.  In general $u$ will not have a
continuous extension to $\partial\D(\Sigma)$ but the direction it
determines does have a $C^{k-1,1}$ extension.  Let $\hat u$ denote a
future--pointing vector field which is proportional to $u$ on
$\stackrel{\circ}{{\cal D}}\!\!  (\Sigma)$ and which extends in a
$C^{k-1,1}$ manner to $\partial\D(\Sigma)$.  (For instance $\hat u$
could be taken to be a vector proportional to $u$ which is unit with
respect to a smooth auxiliary Riemannian metric.)  Now a frame will be
constructed which is convenient for calculations.  A null frame is by
definition either a set of (real)  vector
fields $(l, n, x, y)$ with $g(l, n) = 1$, $g(x, x) = g(y, y) = -1/2$
and all other inner products zero, or the set of vector fields $(l, n,
m\equiv x+ i y)$ derived from the former. Here and elsewhere
$i\equiv\sqrt{-1}$. In the following $l$ and $n$ will be assumed
future pointing and Cauchy horizons will be assumed without loss of
generality to be future horizons.  Thus $l$ and $n$ point from the
globally hyperbolic  region towards the Cauchy horizon; alternatively
we can say that $l$ points away from $\ods$.

\begin{Lemma} \label{1}
  Let $\gamma$ be a null geodesic in $\stackrel{\circ}{{\cal D}}\!\!
  (\Sigma)$ which has    an
  endpoint $p$ on $\partial \D(\Sigma)$.  Then there exist an open
  neighborhood ${\cal V}_0$ of $p$ in $\partial \ds$, an open
  neighborhood ${\cal V}$ of $\gamma$ in $\stackrel{\circ}{{\cal
      D}}\!\! (\Sigma)$ such that $\overline{\cal V}
\ni {\cal V}_0$, a  neighborhood
  $\cal{U}$ of $e$ in $G$ and a smooth null frame $(l, n,
  x, y)$ on ${\cal{V}}$ such that
\begin{enumerate}
\item $l$ is tangent to $\gamma$,\label{i}
\item the vectors $x$ and $y$ are tangent to the group orbits,
\item the frame is invariant in the sense that if $p,q \in {\cal V}$, $g
  \in \cal{U}$ and $\phi_g(p)= q$ then the derivative of $\phi_g$ maps the
  frame at $p$ to the frame at $q$,\label{iii}
\item the Newman--Penrose (NP) coefficients $\epsilon$ and $\kappa$
  vanish,\label{iv}
\item the frame has a $C^{k-1,1}$ extension to ${\cal V}_0$.\label{v}
\end{enumerate}
\end{Lemma}
\proof\ Let $p$ be the endpoint of $\gamma$ on $\partial\D(\Sigma)$
and $l$ a future--pointing tangent vector to $\gamma$ there.  There
exists a neighbourhood $\V_0$ of $p$ in $\partial\D(\Sigma)$ and a
neighbourhood $\cal{U}$ of $e$ in $G$ such that the mapping $g \mapsto
\hat{\group}_g(p)$ of Proposition \ref{P1} is a $C^{k-1,1}$
diffeomorphism of $\cal{U}$ onto $\V_0$.  Extend
$l$ to $\V_0$ by invariance under $G$, i.e. if $q \in \V_0$ let $l(q)
= (\hat{\phi}_{g})_{*} (l(p))$, where $g$ is the unique element of
$\cal{U}$ with $\hat{\group}_g(p) = q$. Here ${\psi}_{*}$
denotes  the tangent map of the map $\psi$.  Let
$\cal{V}$ be the union of the inextendible null geodesics in
$\stackrel{\circ}{{\cal D}}\!\!  (\Sigma)$ through points of $\V_0$
with initial vector $l$.  Extend $l$ to $\cal{V}$ by requiring it to
be the tangent vector to those geodesics at every point.  Now $l$ and
$u$ define a smooth 2--dimensional distribution $\Pi$ in $\cal{V}$.
Let $n$ be the unique future--pointing null vector contained in $\Pi$
and satisfying $g(l,n) = 1$.  Let $x$ and $y$ be smooth vector fields
of length $1/\sqrt{2}$ on $\cal{V}$ which are invariant under $G$ and form an
orthogonal basis of the orthogonal complement of $\Pi$.  The vector
field $n$ clearly has a $C^{k-1,1}$ extension to $\partial\D(\Sigma)$
and $x$ and $y$ can be chosen so that they also extend.  In this way a
frame is obtained which has properties \ref{i}--\ref{iii} and \ref{v}.
There remains some arbitrariness in the choice of $x$ and $y$ and this
will now be used to arrange \ref{iv}.  The fact that $l$ is tangent to
a congruence of affinely parametrized null geodesics implies that
$\kappa$ vanishes and that $\epsilon$ is purely imaginary.

Let $t$ be an affine parameter on $\gamma$.  Then the residual freedom
in $x$ and $y$ is a $t$--dependent rotation.  Call the angle of rotation
$\theta(t)$. Then under a change of basis $\IM\epsilon$, where
$\IM\epsilon$ is the imaginary part of $\epsilon$,  changes by a
constant multiple of ${d\theta/ dt}$.  Hence a rotation of $x$
and $y$ can be chosen so that for the new frame $\epsilon=
0$. \hfill $\Box$

The vector $u$ is a linear combination of $l$ and $n$.  The
normalization conditions which
have been imposed imply that it is of the form ${1\over \sqrt{2}}
(z^{1/2} l + z^{-1/2}n)$ for some non--vanishing smooth\footnote{$z$
can be seen to be smooth by general considerations. Alternatively its
smoothness follows from Proposition \ref{PU2} below.} function $z$
on $\stackrel{\circ}{{\cal D}}\!\!  (\Sigma)$.  As $\partial
\D(\Sigma)$ is approached $u$ becomes parallel to $n$ and hence $z
\rightarrow 0$.  The NP equations which will be needed will now be
written out.  As usual in the NP formalism $l$ and $n$ will be denoted
by $D$ and $\Delta$ when they are thought of as differential operators
on functions.  The NP coefficients are invariant under the group and
so only depend on $t$, the affine parameter along $\gamma$.  The same
is true of $z$. Hence the differential operators associated to $x$
and $y$ annihilate these quantities and when acting on these
quantities $D$ and $\Delta$ are related by $\Delta = zD$.  The
relevant NP equations can be found in \cite[Appendix A]{NewmanTod} or
\cite[Appendix A]{SiklosCMP}, we reproduce them here for the
convenience of the reader.
\begin{eqnarray}
\label{(a)}
D\rho & = & \rho^2 + \sigma\bar\sigma\,,\\
D\sigma & = & (\rho+ \bar{\rho})\sigma + \Psi_0\,,\\
D\tau& = & (\tau+ \bar{\pi}) \rho+ (\bar {\tau} + \pi) \sigma + \Psi_1\,,\\
D\alpha & = & \rho\alpha + \beta\bar{\sigma} + \rho\pi\,,\\
D\beta & = & (\alpha + \pi) \sigma + \bar{\rho}\beta + \Psi_1\,,\\
D\gamma & = & (\tau+ \bar{\pi})\alpha + (\bar{\tau} + \pi)\beta +
\tau\pi + \Psi_2\,,\\
D\lambda & = & (\rho\lambda + \bar{\sigma}\mu) + \pi^2 + (\alpha -
\bar{\beta})\pi\,,\\
D\mu & = & (\bar{\rho}\mu + \sigma\lambda) + \pi\bar{\pi} - \pi
(\bar{\alpha} - \beta) + \Psi_2\,,\\
\label{(i)}
D\nu- zD\pi & = & (\pi + \bar{\tau})\mu + (\bar{\pi} + \tau)\lambda
+ (\gamma - \bar{\gamma})\pi + \Psi_3\,,\\
\label{B2}
zD\Psi_0 & = & (4\gamma - \mu)\Psi_0 - 2(2\tau + \beta)\Psi_1 +
3\sigma \Psi_2
\,,\\ \label{(j)}
zD\lambda & = & - (\mu + \bar{\mu}) \lambda -
(3\gamma-\bar{\gamma})\lambda + (3\alpha +
\bar{\beta} + \pi -\bar{\tau}) \nu - \Psi_4\,,\\
\label{(k)}
0 & = & \rho (\bar{\alpha} + \beta) - \sigma (3\alpha-\bar{\beta})
+ (\rho - \bar{\rho}) \tau-
\Psi_1\,,\\
\label{(l)}
0 & = & (\mu \rho-\lambda\sigma) + \alpha\bar{\alpha} + \beta\bar{\beta}
- 2\alpha\beta + \gamma(\rho-\bar{\rho}) - \Psi_2\,,\\
\label{(m)}
0 & = & (\rho-\bar{\rho})\nu + (\mu-\bar{\mu})\pi + \mu (\alpha +
\bar{\beta}) + \lambda
(\bar{\alpha}-3\beta) - \Psi_3\,,\\
\label{(n)}
-zD\mu & = & (\mu^2 + \lambda\bar{\lambda}) + ( \gamma + \bar{\gamma})
\mu - \nu\pi + (\tau
- 3\beta -\bar{\alpha})\nu\,,\\
-zD\beta & = & (\tau-\bar{\alpha} - \beta)\gamma + \mu \tau- \sigma \nu -
\beta (\gamma - \bar{\gamma} - \mu) + \alpha \bar{\lambda}\,,\\
-zD\sigma & = & (\mu\sigma + \bar{\lambda}\rho) + (\tau + \beta -
\bar{\alpha}) \tau - (3\gamma
- \bar{\gamma})\sigma\,,\\
zD\rho & = & -(\rho\bar{\mu} + \sigma \lambda) + (\bar{\beta} - \alpha -
\bar{\tau})\tau+ (\gamma
+ \bar{\gamma})\rho- \Psi_2\,,\\
\label{(r)}
zD\alpha & = & \rho\nu - (\tau + \beta)\lambda + (\bar{\gamma} - \bar{\mu})
\alpha +(\bar{\beta} - \bar{\tau})\gamma - \Psi_3\,.
\end{eqnarray}
These equations hold on $\ods$ and all the quantities occuring there
are bounded in a (one--sided) neighborhood of $\partial {\cal D}(\Sigma)$.
Using the equations it can be seen that all quantities except possibly
$\Psi_0$ extend continuously to $\partial \D(\Sigma)$.

A central part of this paper is to discuss the possible solutions of
the above set of equations and the first step is to discuss the
solution on $\partial \D(\Sigma)$. By applying the commutator
equations which define the NP coefficients to an arbitrary
$G$--invariant function Siklos \cite{SiklosCMP} obtains the equations
\begin{eqnarray}\label{B3}
Dz & = & - (\gamma + \bar{\gamma})\,,\\ \label{B4}
0 & = & \kappa z + (\bar{\alpha} + \beta - \bar{\pi})\,,\\
0 & = & - \bar{\nu} + (\tau - \bar{\alpha} - \beta) z\,,\\
0 & = & \mu - \bar{\mu} + (\rho - \bar{\rho}) z \,.\label{B6}
\end{eqnarray}
Thus on $\partial D(\Sigma)$ we must have $\pi = \alpha +
\bar{\beta}$, $\nu = 0$ and $\IM\mu = 0$.  He also shows, by
transforming from the null frame to a more conventional orthonormal
frame, that Bianchi class A solutions (which include those of type IX)
satisfy
\begin{eqnarray*}
\mu + \rho z & = & 0\,,\\
\tau - 2\beta - \kappa z & = & 0 \,.
\end{eqnarray*}
Hence in this case $\mu = 0$ and $\tau = 2\beta$ on the horizon.  The
mean extrinsic curvature of the orbits is given by $\sqrt{2} \RE \{(\epsilon -
\bar{\rho}) z^{1/2} + (\mu - \bar{\gamma})z^{-1/2}\}$, where $\Re(x)$
or $\Re x$ denotes the real part of $x$, which for a class A solution
is equal to $-\sqrt{2}\Re \gamma z^{-1/2} + O(z^{1/2})$.  In
\cite{AlanBianchi} it was shown that for Bianchi IX space--times the
mean extrinsic curvature must tend monotonically to infinity at the boundary of
the maximal Cauchy development ({\em cf.\/} Lemma \ref{Alanlemma} below
for a detailed argument) and so it can be concluded that $\gamma +
\bar{\gamma} \not= 0$ on $\partial\D(\Sigma)$.

\begin{Lemma} \label{2} Let $\gamma$
  be a null geodesic as in Lemma \ref{1}.  Then
  there exists a possibly different null geodesic $\bar{\gamma}$ with
  the same endpoint such that on $\partial\D(\Sigma)$ the only
  non--vanishing NP coefficients of the frame associated to
  $\bar{\gamma}$ as in Lemma \ref{1} are $\gamma$ and $\rho$.
  Moreover $\bar{\gamma}\rho+\gamma \bar{\rho} =0$ there.
\end{Lemma}
\proof\ From the above discussion we know that $\nu = 0$, $\mu = 0$,
$\tau = 2\beta$ and $\gamma + \bar{\gamma} \ne 0$ on $\partial
\D(\Sigma)$. A null rotation about
$n$ can be used to set $\sigma$ to zero \cite{SiklosCMP}.  The
geodesic $\bar{\gamma}$ is chosen to have the transformed $l$ as
tangent vector.  The remaining statements follow from the NP equations
 (\ref{(j)})--(\ref{B6}) at
$z=0$ by straightforward algebra.  \hfill $\Box$

\begin{Proposition}\label{PU2}
 Consider the system of equations (\ref{(a)})--(\ref{B2}) and \eq{B3}
for the vector valued function $$f = (\rho,
\sigma, \tau, \alpha, \beta, \gamma, \lambda, \mu, \nu,z, \Psi_0) \in
\C^{\myspace 11} \,,$$
in which
the $\Psi_a$'s, $a = 1, 2, 3$ are determined by
$f$ via eqs \eq{(k)}--\eq{(m)}, and $\pi$ is determined by \eq{B4}
with $\kappa=0$.
Suppose moreover that $\gamma$ and $\mu$ are continuous and that
\be\Re \gamma (0) \not= 0, \qquad \Re\left({4\gamma-\mu\over \gamma +
\bar{\gamma}}\right)(0) > 0\,.
\label{SIK.0}\ee
Assume finally that $z$ is continuous with $z(0)=0$, and $z(x)>0$ for
$x\ne 0$. Then: \begin{enumerate}\item Let $\epsilon > 0$ and let $f
\in L^\infty([0,
\epsilon] ; \C^{\myspace 11}) \cap C^1((0, \epsilon); \C^{\myspace
  11})$ satisfy\,\footnote{For an open set $\Omega$ we use the standard
  notation $C^1(\Omega;X)$ to denote the collection of $X$--valued
  functions on $\Omega$ which are differentiable on $\Omega$ --- no
  uniform bounds are implied; similarly for spaces with higher
  derivative index. On the other hand $C^k(\bar\Omega; X) $,
  respectively $C^\infty(\bar\Omega; X) $ denote the spaces of those
  smooth functions on $\Omega$ the derivatives of which of order less
  than or equal to $k$, respectively of all orders, can be extended by
  continuity to continuous functions on the closure $\bar \Omega$ of
  $\Omega$. $L^\infty(\Omega;X)$ denotes the collection of
Lebesgue--measurable $X$--valued functions on $\Omega$ which are
essentially bounded on $\Omega$.} these equations. Then $f$ can be extended by
continuity to a continuous function on $[0,\epsilon]$, still denoted by
$f$,  such that
$$f \in C^\infty([0, \epsilon]; \C^{\myspace 11})\, .$$
\item Let $\epsilon > 0$ and let
$f_1, f_2 \in C^0([0, \epsilon]; \C^{\myspace 11})
\cap C^1((0, \epsilon), \C ^{\myspace 11})$ be two solutions of those
equations satisfying our previous hypotheses, with
$$f_1(0) = f_2(0)\,.$$
Then
$$f_1 \equiv f_2\, .$$
\end{enumerate}
\end{Proposition}

\Proof\ Writing $f$ as $(g,\Psi_0)$ and using coordinates in which $D
= -{d\over dx}$, equations
\eq{(a)}--\eq{B2} together with  eq. \eq{B3}
are of the form
\begin{equation}
x{d\over dx}\left({g \atop \Psi_0}\right) =
\left({xB(f)\atop C(f_,x)}\right)\, ,
\label{SIK.1}\ee
where
\begin{equation}
C(f,x) = {x\over z}E(f)\, , \label{SIK.1.1}\ee
with $E(f)$ given by the right hand side of eq. \eq{B2}, and $B(f)$
given by the right hand side of eqs. \eq{(a)}--\eq{(i)} and \eq{B3}.
Note that by the first part of (\ref{SIK.0}) and by (\ref{B3}) we can
choose $\delta$ small enough so that $x/z$ is bounded on $[0,\delta]$.
{}From our hypotheses on $f$ it follows immediately
that $dg/dx \in L^\infty([0, \epsilon]; \C^{\myspace 10})$, so that
$g$ can be uniquely extended to a Lipschitz continuous function on
$[0,\epsilon]$, still denoted by $g$.
{}From \eq{B3} we then have $\lim_{x\to 0}x/z =
1/(2\RE\gamma)(0)$.   The
equation for $\Psi_0$ can be rewritten in the form
\begin{equation}
x{d\over dx} (\Psi_0(x)-\Psi_0(0)) + \chi(x)(\Psi_0(x)-\Psi_0(0)) = F(x)\,,
\label{SIK.2}
\end{equation}
with
$$
\chi(x) = -{(4\gamma - \mu)x\over z}\, ,\qquad \Re\chi(0) = \Re\{{
4\gamma-\mu\over
2\RE\gamma} \} > 0\,.
$$
Here the constant $\Psi_0(0)$ has been chosen so that $F(0)=0$; that
this can always be achieved follows from our hypotheses which
guarantee that $\chi(0)\ne 0$.
Integrating \eq{SIK.2} it follows that there exists a constant $C$
such that
\begin{equation}
\label{newintegral}
\Psi_0(x)-\Psi_0(0) = \left[C-\int_x^\epsilon
\frac{F(s)}{s} \,\exp^{\left(-\int_s^\epsilon
\frac{\chi(t)}{t}dt\right)}ds\right]\exp^{\left(\int_x^\epsilon
\frac{\chi(t)}{t}dt\right)} \,.
\end{equation}
Clearly $|\Psi_0|$ will blow up as $x^{-\Re \chi (0)}$ at $x=0$ unless
the constant $C$ is appropriately chosen. Let us set
$$
\Xi(x) = \exp\left( \int^x_0 {\chi(0)-\chi(s)\over s} ds\right) .
$$
{}From (\ref{newintegral}) and from our hypothesis $\Psi_0\in
L^\infty([0,\epsilon],\C)$ it therefore follows
\begin{equation}
\Psi_0(x)-\Psi_0(0) = \left[\int_0^x {{F(s)}\over \Xi(s)} s^{\chi(0)-1}
ds\right] x^{-\chi(0)} \Xi(x)\, , \label{SIK.5}
\end{equation}
which immediately gives
$$ |\Psi_0(x) - \Psi_0(0)| \leq Cx .  $$ An
easy bootstrap of (\ref{SIK.5}) and of (\ref{SIK.1}) establishes point
1.  To prove point 2, let $f_a=(g_a,\Psi^a_0)$, $a=1,2$, be two
solutions and define
\begin{eqnarray*}
 & \Delta g(x) = g_1(x) - g_2(x)\, ,&\\& \Delta\Psi(x) = \Psi^1_0(x)
  -\Psi^2_0(x)\, ,&\\& H  = x^{-1} \langle\Delta g, \overline{\Delta
    g}\rangle + x^{-1/2} |\Delta\Psi|^2\, .
\end{eqnarray*}
Here a bar denotes complex conjugation, and $\langle\ ,\ \rangle $ is
the standard scalar product on $\R^{\myspace 10}$. We have
  \begin{equation}
{dH\over dx} = -x^{-2} \langle\Delta g, \overline{\Delta
    g}\rangle -{1\over 2} x^{-3/2} |\Delta\Psi|^2+ G \, ,
 \label{SIK.5.1}
\end{equation}
where by $G$ we have denoted the sum of all the remaining terms which
arise by differentiation of $H$. One of those terms, {\em e.g.}, will
be
$$ x^{-3/2} \left\{\Delta\bar\Psi \left[ {x\over z_1} E(f_1) - {x\over z_2}
E(f_2) \right] + c.c.\right\}\,, $$
with $E(f)$ as in \eq{SIK.1.1}, and where $c.c.$ denotes the complex conjugate
term. This can be rewritten as
\begin{equation}
 x^{-3/2} \left\{\Delta\bar\Psi \left[ {x\Delta
z \over z_1z_2} E(f_1) + {x\over z_2} \left(E(f_1)- E(f_2)\right)
\right] + c.c.\right\}\,.
\label{SIK.5.11}
\end{equation}
 In what follows the letter $C$ will denote a large constant which
may vary from expression to expression. Now by eq.\ \eq{B3} for $x$
small enough we have $z_a\ge C^{-1}x$, $a=1,2$. Moreover \eq{SIK.1}
and $\Psi^1_0\in C^1([0,\epsilon];\C)$ imply that $|E(f_1)|\le Cx$.
Consequently, for any $\alpha >0$ we have
 \begin{eqnarray}
\nonumber x^{-3/2} |\Delta\bar\Psi  {x\Delta
z \over z_1z_2} E(f_1)| & \le & C x^{-3/2} |\Delta \Psi|\, |\Delta z|
\\
\label{SIK.5.2} &
\le & {C\over \alpha} x^{-3/2}|\Delta z|^2 + {C\alpha\over 4} x^{-3/2}|\Delta
\Psi|^2\,.
\end{eqnarray}
The second term in the right--hand--side of this equation will be dominated
by the second term in \eq{SIK.5.1} if we choose $\alpha$ small enough.
Then we can find $\epsilon_1>0$ such that for $x\in(0,\epsilon_1)$ the
first term in \eq{SIK.5.2} will be dominated by the first term in
\eq{SIK.5.1}.

To analyze the remaining terms in \eq{SIK.5.11}, let us write the expression
$E(f)$ in eq.\  \eq{SIK.1.1} in the form
$$
E(f) = -\chi(g)(\Psi_0 -\Psi_0(0))+ F(g)\,.
$$
We then have
\begin{eqnarray*}
& \Delta\bar\Psi   \left(E(f_1)- E(f_2)\right)
 + c.c.  =  \nonumber & \\
&  = \Delta\bar\Psi
\left(-\chi(g_1)(\Psi_0^1-\Psi_0(0))+\chi(g_2)(\Psi_0^2-\Psi_0(0))+ F(g_1)-
F(g_2)\right)+ c.c.\nonumber& \\
 & =  \left\{-\chi(g_1)|\Delta\Psi|^2 +\Delta\bar\Psi\left[(
\chi(g_2)-\chi(g_1))(\Psi_0^2-\Psi_0(0))+ F(g_1)-
F(g_2)\right]\right\}+ c.c.&
\nonumber \\
 & = -2\Re\chi(g_1)\ |\Delta\Psi|^2 + \left\{\Delta\bar\Psi\left(
\chi(g_2)-\chi(g_1))\Psi_0^2+ F(g_1)-
F(g_2)\right)+ c.c.\right\}\nonumber
\,. &
\nonumber
\end{eqnarray*}
The first term in this expression is negative for $x$ small enough,
and can be dropped, while the remaining terms can be estimated as in
the previous calculation.
Proceeding similarly with the remaining terms which
occur in $G$ in \eq{SIK.5.1} we find that there exists
$0 < \delta \leq \epsilon$ such that for $x \in (0, \delta)$ we have
$${dH\over dx} \leq 0 .$$ As $H$ is continuous on $[0,\delta]$ with
$H(0) = 0$ we obtain $$ H\equiv
0\Rightarrow f^1 \equiv f^2$$ on $[0, \delta]$.  $f^1$ $\equiv f^2$ on
$[0, \epsilon]$ follows now from standard uniqueness results for
(non--singular) ODE's.\hfill $\Box$

It has now been shown that in a Bianchi IX spacetime with a Cauchy
horizon the spin coefficients in a certain frame take a special form
on the horizon.  It has also been shown that the spin coefficients are
determined everywhere by their values on the horizon.  The spin
coefficients of certain frames in the Taub--NUT spacetimes will be
calculated and compared with the restrictions previously obtained.
The Taub--NUT metrics can be written (perhaps locally) in the form
\cite{Misner}
\begin{eqnarray}
& U^{-1}dt^2 -[(2L)^2U\sigma^2_1 + (t^2 + L^2)(\sigma^2_2 +
\sigma^2_3)]\,, \label{MBianchi}
& \\ &
U(t) = -1 + {2(mt + L^2)\over
  t^2 + L^2}\,. \label{UBianchi}
\end{eqnarray}
where $\sigma_1$, $\sigma_2$ and $\sigma_3$ are, say, left invariant
 one--forms on $SU(2)$.  The constants $L$ and $m$ are real numbers
 with $L > 0$.  Let $X_1$, $X_2$, $X_3$ be dual basis vectors to
 $\sigma_1$, $\sigma_2$ and $\sigma_3$.  They have the commutation
 relations $[X_i, X_j] = {\epsilon^k}_{ij} X_k$. Two null vectors can
 be defined by
\begin{eqnarray}\label{ldef}
& l = {1 \over \sqrt{2}} \left({\partial\over \partial t} \pm {1\over
  2LU} X_1\right)\,, &
\\ \label{ndef}&
n = {1 \over \sqrt{2}} \left(U{\partial\over
\partial t} \mp {1\over 2L} X_1\right)\,. &
\end{eqnarray}
They can be completed to a basis by $x = {1\over
  \sqrt{2(t^2+L^2)}}X_2$ and $y ={1\over\sqrt{2(t^2+L^2)}} X_3$.  The
non--vanishing spin coefficients of the frame $(l, n, x, y)$ are given
by
\begin{eqnarray*}
& \rho  =  {1\over \sqrt{2}} \left(- {t\over t^2+L^2} \pm {L\over
  t^2+L^2}i\right) \,,
&\\ &
\mu  =  {1\over \sqrt{2}} \left( {Ut\over
  t^2+L^2} \mp {LU\over t^2+L^2}i\right) \,,
&\\ &
\epsilon  =  {1\over 2
  \sqrt{2}} \left(\pm {L\over t^2+L^2} \mp {1\over 2LU}i\right) \,,
&\\ &
\gamma  =  {1\over 2 \sqrt{2}} \left[-\dot U + \left(\mp {LU\over t^2+L^2}
\pm {1\over 2L}\right)i\right]\,. &
\end{eqnarray*}
Here and elsewhere a dot over a quantity denotes a time
derivative.
Since $\epsilon  \not= 0$ this is not a frame of the type used by Siklos and
constructed in Lemma \ref{1}.  However, as in the proof of that lemma,
$\epsilon$ can be set to zero by doing a rotation of $x$ and $y$.
This rotation leaves the other spin coefficients except $\gamma$
unchanged.  In the new frame
$$\gamma = {1 \over 2\sqrt{2}}\left[ - \dot{U} + \left(
\mp {2LU\over t^2 + L^2}
\pm {1\over L} \right)i\right]\,.$$
On the future horizon $\dot{U} = -{4t(m^2 +L^2)\over
(t^2+L^2)^2} $ and it can be checked directly that
$\bar{\gamma}\rho + \gamma\bar{\rho} = 0$ there.

In Lemma \ref{2} it was shown that on the Cauchy horizon of a Bianchi
IX space--time $\RE(\gamma\bar{\rho}) = 0$.  The restriction $\gamma +
\bar{\gamma} \not= 0$ was also obtained.  The fact that a future
horizon is considered means\footnote{Let us elaborate slightly on this
point. By choice the function $z$ has been chosen to be positive on
$\ods$. Now if we choose $l$ to be pointing {\em away} from $\ods$,
then \eq{B3} implies that $\Re \gamma $ must be nonnegative on the
Cauchy horizon. Alternatively we could ask that $l$  be future
pointing -- in that case we have  $ \Re \gamma \ge 0 $ on a future
Cauchy horizon and  $ \Re \gamma \le 0 $ on a past one.}
that in fact $\gamma + \bar{\gamma} > 0$.  There is a further
restriction which follows from the criteria given\footnote{The reader should be
  warned of \label{misprintsfootnote} misprints in eq.\ (3.2) of
\cite{SiklosCMP}.} by Siklos
\cite[Section 3]{SiklosCMP}, namely that $(\IM\gamma)(\IM\rho) < 0$.
Define
\begin{equation}
 S_1 = \{(\rho, \gamma) \in \C^{\myspace 2} :
 \RE (\gamma\bar{\rho}) = 0, \gamma +
\bar{\gamma} > 0, (\IM\gamma)(\IM\rho) < 0 \} \,.
\label{defS1}
\end{equation}
The family of spin coefficients arising from Taub--NUT space--times
described above can be extended by using a constant boost of $l$ and
$n$.  Under a boost $\rho$ and $\gamma$ transform according to
$\rho\rightarrow A\rho$, $\gamma \rightarrow A^{-1}\gamma$, where $A$
is an arbitrary positive real number.  A calculation reveals that the
resulting family of spin coefficients depending on the three
parameters $L$, $m$ and $A$ precisely exhausts the set
$S_1$.
(The two components of $S_1$
defined by the sign of $\IM\gamma$ correspond to the two possible
signs in the definition \eq{ldef}--\eq{ndef} of $l$ and $n$.)
Thus given a Bianchi IX space--time with a horizon there exists a
Taub--NUT space--time and a null frame there which has the same spin
coefficients as a frame in the given space--time.

 \begin{Lemma}\label{3} Let $(M, g)$ be a vacuum Bianchi IX space--time with
   partial Cauchy surface $\Sigma$, as defined at the beginning of
   this section and let $p$ be a point of $\partial {\cal D}(\Sigma)$.
   Then there exists a Taub--NUT space--time $(M', g')$ and an
   isometry $\phi_p$ of a neighbourhood $\cW_p$ in
   $\overline{{{\cal D}} (\Sigma)}$ of    $p$ onto a
   subset of $(M', g')$.
\end{Lemma}
\proof\ It has already been shown that there exist a null geodesic
$\gamma$ with endpoint $p$, a Taub--NUT space--time $(M', g')$ and a
null geodesic $\gamma'$ in $M'$ with endpoint $p'$ such that the
frames associated to $\gamma$ and $\gamma'$ as in Lemma \ref{1} have
the same spin coefficients.  If $q$ belongs to the neighbourhood
$\cV_0$ of $p$, let $g$ be the unique element of ${\cal U}$ with
$\hat{\group}_g(p) = q$ and define $\phi(q) = \phi'_g(p')$, where
$\phi'$ denotes the action of $G = SU(2)$ or $G = SO(3)$ on the
Taub--NUT space--time.  The mapping $\phi$ is a diffeomorphism of
$\cV_0$ onto a neighbourhood of $p'$ in the
horizon of the Taub--NUT solution.  The frame on the Taub--NUT
space--time is not uniquely defined by $\gamma$ and $\gamma'$ alone.
There remains the freedom
to do Lorentz transformations which preserve $n$.
However the requirement that the NP coefficients be identical fixes
the two null vectors in each case uniquely.
Thus $\phi_{*,p}(l) = l'$ and $\phi_{*,p}(n) = n'$. Here
$\phi_{*,p}$ denotes the tangent map at $p$ associated to the map
$\phi$. The vectors $\phi_{*,p}(x)$ and $\phi_{*,p}(y)$ may differ
from $x'$ and $y'$ by a rotation.  However this rotation does not
change the NP coefficients and so we may define $x' = \phi_{*,p}(x)$
and $y' = \phi_{*,p}(y)$. Now the frame at $p'$ agrees with the image
under $\phi$ of the frame at $p$.  Using the group invariance shows
that the frames agree on $\partial {\cal D}(\Sigma)$.  The mapping
$\phi$ will now be extended to a neighbourhood of $p$.  Let $q$ be a
point which lies on a null geodesic with endpoint on $\cV_0$ and
has tangent vector $l$ there.  Let $r$ be the endpoint of this geodesic
and $t$ the affine distance from $r$ to $q$.  Define $\phi(q)$ to be
the point obtained by following the geodesic through $\phi(r)$ with
tangent vector $\l'$ there an affine distance $t$ into the
past. To complete
the proof of the lemma it suffices to show that $\phi$ is an isometry
and to do that it is enough to show that $\phi$ maps the frame $(l, n,
x, y)$ to the frame $(l', n', x', y')$
everywhere (note that at this stage we know only that this is true on
a subset of the horizon).  To see that this is true, note first that
the construction
of $\phi$ ensures that $\phi_*l = l'$ wherever defined. The definition of
the NP coefficients allows the Lie derivatives of $n$, $x$ and $y$ to
be expressed as linear combinations of $l$, $n$, $x$ and $y$, where
the coefficients are expressions involving the NP coefficients. Of
course the same is true of the Lie derivatives of $n', x'$ and $y'$.
Because the spin coefficients are the same for both frames this means
that the components of $\phi_*n$, $\phi_*x$, $\phi_*y$ and those of
$n', x', y'$ satisfy the same system of linear ordinary differential
equations.  Since $\phi_*n = n'$, $\phi_*x = x'$ and $\phi_*y = y'$ on
$\partial {\cal D}(\Sigma)$ they also have the same initial data at
some point. Hence $\phi_*n = n'$, $\phi_*x = x'$ and $\phi_*y = y'$ on
a neighbourhood of $p$.\hfill $\Box$

{\bf Proof of Theorem \ref{T1}:} It follows from the construction of
the local isometry $\phi$ in the proof of Lemma \ref{3} that $\phi$
maps the intersection of each hypersurface of homogeneity with ${\cal
W}_p$ into a hypersurface of constant $t$ in the Taub--NUT
spacetime. Here $t$ is as in eq.\ \eq{MBianchi}. Let $(\ga,K)$ be the
initial data which $(M,g)$ induces on the hypersurface of homogeneity
$\Si_q$ containing a point $q\in {\cal W}_p \cap \cD(\Si)$ and let
$(\ga',K')$ be the initial data on the corresponding hypersurface
$\Si_q'$ of constant $t$ in Taub--NUT spacetime. Then $\phi_*\ga=\ga'$
and $\phi_* K=K'$, wherever the left hand sides of these equations are
defined. Now if $\Si_q'$ is simply connected, it can be identified
naturally with $SU(2)$, and then $\ga',K'$ are left-invariant tensors
there. It can be assumed without loss of generality that $\phi(q)$ is
identified with $e$. If $\Si_q$ is simply connected then there is a
diffeomorphism $\phi_1:\Si_q\to SU(2)$ with $\phi_1(q)=e$ such that
$(\phi_1)_*\ga$
 and
$(\phi_1)_*K$ are left-invariant. Now $\phi_1\circ\phi$ is a mapping
$SU(2)\to SU(2)$ such that $(\phi_1\circ\phi)_*\ga'=\phi_1{_*}\ga$ on a
neighbourhood of the identity. Since both of these tensors are left
invariant they must be equal everywhere. A similar argument applies
to $K$. Hence $\phi$ maps $(\Si_q,\ga,K)$ onto $(\Si_q',\ga',K')$ and
their maximal globally hyperbolic developments must be isometric. It
follows that $\cD(\Si)$ is isometric to (the globally hyperbolic part
of) a Taub--NUT spacetime. If $\Si_q$ is not simply connected we get
the statement that $\cD(\Si)$ is isometric to a quotient of a Taub--NUT
spacetime by a discrete group of isometries, which are  the Taub--NUT
spacetimes with lens space spatial topology.

To finish the proof of Theorem \ref{T1} we need to prove that generic
Bianchi IX initial data on $L(p,1)$, $p=1,2$, are {\em not} of Taub--NUT
type. The result is well known on $L(1,1)=S^3$, {\em cf.\ e.g.\/}
\cite{SCC} or \cite{RyanShepley}. For $p>1$ recall that $L(p,1)$ is
defined as $S^3/{\Z}_p$, for a suitable action of ${\Z}_p$ on
$S^3$. We have the following:

\begin{Proposition}\label{Ptop}
1. All Bianchi IX initial data on $S^3$ pass to the quotient
$S^3/{\Bbb Z}_2\approx L(2,1)$.

2. For $p>2$ all (globally) homogeneous Bianchi IX initial data on
$L(p,1)$ are of Taub--NUT type.
\end{Proposition}

{\bf Remark:} Let us emphasize that there exist non--Taub--NUT Bianchi
IX metrics on $L(p,1)$ for any $p>1$; {\em cf.\/} the discussion in
Section \ref{topology} below. These will, however,
be only {\em locally} homogeneous for $p>2$ if they are not of
Taub--NUT type.

{\bf Proof:} Using the Euler angle parametrization of $S^3$ ({\em cf.\
e.g.\/} eq.\ (\ref{Eulermetric}) below) the relevant $\Z_p$ action
consists of translations in $\mu$ and $\varphi$:
\begin{equation}
\mu\to\mu+
\frac{2\pi}{p}, \qquad
\varphi\to\varphi+\frac{2\pi}{p}\,.\label{ptr}
\end{equation}
Let $\phi$ denote the corresponding map from $S^3$ into itself.
Consider\footnote{To obtain the standard form of the $SU(2)$
commutation relations the vectors $X_1$ and $X_2$ given in
\cite{ChImaxTaubNUT} should be interchanged.}
 the Killing vectors $X_a$, $a=1,2,3$ given by equations
{(17a)}--{(17c)} of \cite{ChImaxTaubNUT}. Let $\si^b$, $b=1,2,3$, be
the corresponding dual forms, $\si^a(X_b)=\de_b^a$.  From (\ref{ptr})
one immediately obtains
\begin{equation}
\phi^* \left[
\begin{array}{c}
\si^1 \\
\si^2 \\
\si^3
\end{array}
\right]
= \left[
\begin{array}{ccc}
\cos\frac{2\pi}{p} & -\sin\frac{2\pi}{p} & 0 \\
\sin\frac{2\pi}{p} & \cos\frac{2\pi}{p} & 0 \\
0 & 0 & 1
\end{array}
\right]
\left[
\begin{array}{c}
\si^1 \\
\si^2 \\
\si^3
\end{array}
\right]\,.
\label{str}
\end{equation}
Any $SU(2)$ invariant two-covariant symmetric tensor $f$ on $S^3$ can
be written in the form
\begin{equation}
\label{fmetric}
f=f_{ij}\si^i\otimes\si^j
\end{equation}
where $f_{ij}$ is a symmetric $3$ by $3$ constant coefficients matrix.
 It easily follows from (\ref{str})  that
 1) if  $p=2$, then $f$ passes to the quotient if and only if
$f_{13}=f_{23}=0$,
 2) if $p>2$, then $f$ passes to the quotient if and only if $f$ is
diagonal with $f_{11}=f_{22}$.
 Our claims follow directly from the above. \hfill$\Box$
\def\f#1#2{{\textstyle{#1\over #2}}}
\section{Some remarks on local vs. global degrees of freedom}
\label{topology}
The issue of strong cosmic censorship for spatially locally
homogeneous spacetimes of a given Bianchi type cannot in general be
decided by an analysis of the universal covering spacetimes alone. If
Cauchy horizons could be ruled out for all the covering spacetimes of
a given Bianchi type then there would be nothing left to be
done. However the typical situation is that Cauchy horizons can only
be ruled out for almost all of the covering spacetimes and that there
are exceptional cases where a Cauchy horizon does exist. A given
choice of the topology of a compactification may be incompatible with
almost all covering spacetimes but compatible with the exceptional
geometries. This will now be illustrated by the example of vacuum
Bianchi I spacetimes. In that case the three--metric on the initial
surface in the covering spacetime may be assumed without loss of
generality to be the standard flat metric on $\R^3$. Compactifications
are then obtained by factoring out by discrete subgroups of the
Euclidean group which leave the second fundamental form invariant. In
$\R^3$ the tangent space at any point can be naturally identified with
the tangent space at the origin. This identification leads to a
homomorphism from the Euclidean group to the rotation group $O(3)$. By
homogeneity the second fundamental form is constant when its values at
different points are compared using the natural identification. Hence
the question of which geometries can be compactified reduces to the
following question. Given a discrete subgroup $H$ of the Euclidean
group which defines a compactification, let $\bar H$ be the corresponding
subgroup of $O(3)$. Which constant symmetric tensors $K$ on $\R^3$ are
invariant under $\bar H$? It must also be remembered that only a $K$ which
satisfies the Hamiltonian constraint is of interest in the present context.

There are three cases to be considered, according to whether the
number of distinct eigenvalues of $K$ is one, two or three.  In
the first case the Hamiltonian constraint implies that $K=0$. Any
group $H$ is allowed. This is flat space--time
identified in a way which is compatible with a timelike Killing
vector. The resulting maximal globally hyperbolic developments of the
initial data are static and geodesically
complete. In the second case the Hamiltonian constraint implies that
the eigenvalues of $K$ must be of the form $(\lambda, 0, 0)$ or
$(\f23\lambda,\f23\lambda, -\f13\lambda)$.  These correspond to the
$(1,0,0)$ and $(\f23,\f23,-\f13)$ Kasner solutions respectively. The
$(1,0,0)$ Kasner solutions admit a Cauchy horizon and in fact these
are the only choices of $K$ compatible with a Cauchy horizon. In
particular the third case is not
compatible with a Cauchy horizon. The possible groups $H$ which lead
to a compact orientable quotient manifold have been listed by Koike,
Tanimoto and Hosoya in
\cite{KoikeTanimotoHosoya}, and we shall consider their list one by
one. The first type on their list is the
torus, where the group $\bar H$ is trivial. There all three cases are
possible for $K$ and cosmic censorship holds in the sense that
there is no Cauchy horizon for a generic metric with Bianchi I
symmetry. The second type, where the only non--trival element of $\bar H$
is a rotation through $\pi$ is similar. In the next three types $\bar H$
consists of rotations about a given axis including ones through angles
other than $\pi$. Hence only the first and second cases are possible
for $K$. Thus there are two classes of solutions without Cauchy
horizons and one with Cauchy horizons. Since the class with Cauchy
horizons is not in any obvious sense smaller than the set of solutions
without horizons, it seems reasonable to say that cosmic censorship
fails in this case. Finally, in the sixth type $\bar H$ contains rotations
about distinct axes so that only the first case is possible for
$K$. For this topology all Bianchi I spacetimes are geodesically
complete and cosmic censorship holds.

It should be emphasized that the various cases discussed
above arise because the isometries used to identify the
universal covering space did {\em not} form a subgroup of the Bianchi
group considered. Indeed, consider a simply connected Bianchi group
and let $\Gamma$ be any discrete subgroup thereof. Let $g$ be any
left--invariant metric on $G$, then $g$ passes to the manifold
obtained by dividing $G$ by the left action of $\Gamma$ on $G$.
 The resulting metric will, however, only be {\em locally}
homogeneous in general. This follows from the fact that the Killing
vector fields of the resulting metric on the group manifold are generated by
left
translations of $G$ on itself, which are not globally defined on the
coset space in general.  This is {\em e.g.\/} what happens in the case
of Bianchi IX's on higher $p$ lens spaces; here the $\Z_p$ action
comes from an action of $G$ on itself so that any metric passes to an
appropriate quotient to yield a locally homogeneous metric.

To close this section, let us point out that even in the case of a
solution with a Cauchy horizon and a group which can be used to make
identifications in the globally hyperbolic region, the Cauchy horizon
may not survive the identifications. Consider the chronological future
of the origin in Minkowski spacetime. The group corresponding to
Bianchi type V can be made to act on this region so that the orbits
are the hyperboloids of constant Lorentzian distance from the
origin. In this way this part of Minkowski spacetime can be regarded
as a Bianchi V spacetime. This region can be factored out by the
action of a discrete subgroup of the Lorentz group so as to become
spatially compact. However, it follows from Proposition
\ref{flatLorenz} below that the resulting spacetime
cannot be extended through a Cauchy horizon.

\section{Spatially locally homogeneous 
  space--times with Cauchy horizons}\label{LH} In this section we
shall show how to generalize the results of Section \ref{BIX} to the
spatially locally homogeneous case.  Let $(\Sigma, \gamma, K)$ be a
vacuum initial data set, let $\hat\Sigma$ be the universal cover of
$\Sigma$ and let $\hat\gamma$, $\hat K$ be the lifts of $\gamma$ and
$K$ to $\hat\Sigma$.
We shall say that $(\Sigma,
\gamma, K)$ are locally
homogeneous if there exists a group $G$ acting transitively on
$\hat\Sigma$ such that its action leaves $\hat\gamma$ and $\hat K$
invariant.
{We could similarly define a Riemannian manifold
$(\Sigma,\gamma)$ to be locally homogeneous by dropping $K$ in our
definition above. This apparently differs from the standard definition
given {\em e.g.\/} in \cite{Singer}, where local homogeneity is
defined by postulating existence of an appropriate set of locally
defined isometries. When $\Sigma$ is complete a theorem of Singer
\cite{Singer} shows that these two definitions are actually
equivalent. The approach using locally defined isometries seems to be
somewhat more natural, as it does not require the introduction of
universal covers, etc.   We have chosen the above definition of local
homogeneity
of initial data since no analogue of Singer's theorem including $K$ is
known. } The definition which makes
use of the universal cover turns out to be very
convenient.

Let us start with a brief discussion of the Kantowski--Sachs symmetry
type. These models correspond to the $S^2\times \R$ model discussed in
Section 2.4.8 of \cite{KoikeTanimotoHosoya}. The corresponding initial
data surfaces are obtained by taking compact quotients of $S^2\times
\R$, with the corresponding initial data on $S^2\times\R$ being
invariant under the obvious action of $SO(3)\times\R$ on
$S^2\times\R$. It is well known and in any case easily seen that the
latter initial data can be obtained as data induced on a cylinder
$r=\mbox{const}<2m$ in an extended Schwarzschild space--time.
Similarly the maximal globally hyperbolic development of such data can
be obtained by taking quotients of the $r<2m$ region in the extended
Schwarzschild space--time. It follows that there is always a curvature
singularity in one time direction ($r\to 0$). Moreover when the
initial data surface is compact, then the space--time can always be
extended across a compact Cauchy horizon in the other time direction,
$r\to 2m$. This shows that SCC ``half--fails'' in this class of
space--times. Note, however, that the local algebra of Killing vectors
is always four--dimensional, which establishes the Kantowski--Sachs
part of Theorem \ref{T3}.

To analyze the Bianchi symmetry type, we shall need the following
result (it should be emphasized that compactness of $\Sigma$ is not
assumed here):

\begin{Theorem} \label{TLH.1} Let $(\Sigma, \gamma, K)$ be locally homogeneous
  vacuum initial data, with $(\Sigma,\gamma)$ --- geodesically
complete, and let $(M,g)$ be the maximal globally hyperbolic
development thereof.  Then there exists on $M$ a time function $t \in
C^\infty(M)$ with range $(-\infty, 0)$ the level sets of which are
locally homogeneous Cauchy surfaces.
\end{Theorem}

{\bf Remark:}
 Theorem \ref{TLH.1} allows us to define the notion of  {\em hypersurfaces
of homogeneity}, as the level sets of the function $t$ given there.

\proof\ Let $\hat{\Sigma}$ be the universal cover of $\Sigma$
with projection map $\pi$, set $\hat{\gamma} = \pi^*\gamma$, $\hat{K}
= \pi^*K$, let $G$ be the group appearing in the definition of local
homogeneity,
let $(\hat{M}, \hat{g})$ be the maximal globally hyperbolic
development of $(\hat{\Sigma}, \hat{\gamma},
\hat{K})$.
Recall also that $\pi_1(\Sigma)$ acts on $\hat{\Sigma}$, and
for $g\in \pi_1(\Sigma)$ the action $\pi_g$ thereof satisfies
$\pi\circ \pi_g = \pi$.  We have $$\pi_g^*\hat{\gamma} =
\pi^*_g\pi^*\gamma = (\pi \circ \pi_g)^* \gamma =
\hat{\gamma} , $$ similarly  for $g
\in \pi_1(\Sigma)$ one has $\pi^*_g\hat{K} = \hat{K}$.  By Theorem
2.1.4 of \cite{SCC} there exist actions
$\hat{\phi}_g$ of $G$ and $\hat{\pi}_g$ of $ \pi_1(\Sigma)$ on $(\hat{M},
\hat{g})$ by isometries such that $\hat{\phi}_g|\hat{\Sigma} =
\phi_g$, $\hat{\pi}_g|\hat{\Sigma} = \pi_g$.

Let $p\in \hat{\Sigma}$ and let $\hat{\Gamma} = \{p(s)\}_{s\in I}$ be a
maximally extended affinely parametrized timelike geodesic normal to
$\hat{\Sigma}$ such that $p(0) = p$.  Here $I$ is some open connected
subset of $\Bbb{R}$.
Define a family of surfaces $\hat{\Sigma}_s$, $s \in I$ by $$
\hat{\Sigma}_s = \{\phi_g(p(s)), g \in G\}\,.
$$ It is clearly seen that the $\hat{\Sigma}_s$'s cover $\hat{M}$ and
that the function defined as $t_1(p) = s $ for $p \in \hat{\Sigma}_s $
is a smooth time function on $\hat{M}$.  Let $f : I \rightarrow
(-\infty, 0)$ be any diffeomorphism, set $\hat{t} = f\circ t_1$.  The
function $\hat{t}$ can be used to induce a diffeomorphism $$ \hat{M}
\approx \hat{\Sigma} \times (-\infty, 0)$$ in a standard way.  Define
now $$M = \hat{M}/\pi_1(\Sigma) .$$ Note that for $g \in
\pi_1(\Sigma)$ we have $\hat{t} \circ \pi_g = \hat t$ so that
$\hat{t}$ passes to the quotient and defines a time--function $t \in
C^\infty(M)$.  In particular we have $$M\approx \Sigma \times
(-\infty, 0)\,.$$ We wish to show that $M$ is globally hyperbolic.
Indeed, $M$ is stably causal since it possesses a time function (\cfeg
\cite{Waldbook}).  Moreover $\Sigma \approx \Sigma \times \{t_0\}
\subset M$ is acausal as it is a level set of $t$.  Let $\Gamma$ be a
causal curve in $M$, let $p \in \Gamma$, choose $\hat{p} \in
\pi^{-1}(\Gamma)$ and let $\hat{\Gamma}$ be the connected component of
$\pi^{-1}(\Gamma)$ through $\hat p$. By global hyperbolicity of
$\hat{M}$ the curve $\hat{\Gamma}$ can be extended if necessary to a
curve $\hat{\Gamma}'$ such that $\hat{\Gamma}' \cap \{\hat{\Sigma}\}
\not= \emptyset$, setting $\Gamma' = \pi(\hat{\Gamma}')$ one obtains
an extension of $\Gamma$ which intersects $\Sigma$, and global
hyperbolicity follows.

It remains to show that $M$ is maximal globally hyperbolic.  While one
can give  a general argument using properties of maximal globally
hyperbolic developments and universal covers, we have found it useful
to present the alternative proof that follows, as the arguments below
give some information about the global structure of
$(M,g)$.

The following result has been essentially proved in
\cite{AlanBianchi}:

\begin{Lemma}\label{Alanlemma} Let $(M,g)$ be a vacuum spacetime which
  is the maximal globally hyperbolic development of (locally)
  homogeneous initial data. Let $\tau$ be a coordinate which is constant
  on the hypersurfaces of homogeneity and which coincides with a
  proper time parameter when restricted to geodesics orthogonal to
  these hypersurfaces.  Denote the mean extrinsic curvature of the hypersurface
  labelled by $\tau$ by $H(\tau)$.  Then $H$ is monotone and:
\begin{enumerate}
\item For Bianchi IX or Kantowski--Sachs symmetry
  type$^{\ref{KSfootnote}}$ in each time direction $|H|$ tends to
  infinity in finite time.
\item For the remaining symmetry types, $H$ runs from $-\infty$ to $0$
  (changing time orientation if necessary). The spacetime exists
  globally in $\tau$ in that time direction in which $|H|$ tends to zero,
  say $\tau \in (-\infty,\tau_+)$, for some\footnote{The case $\tau_+={\infty}$
    can occur only if $(M,g)$ is obtained by some identifications from
    Minkowski space--time.}  $\tau_+\in\R\cup{\infty}$, and
  $\lim_{\tau\to-\infty}H=0$.
\end{enumerate}
\end{Lemma}
\proof\ To see this it suffices to note the following facts. Firstly,
the results in \cite{AlanBianchi} are stated for the Vlasov-Einstein
system, where the matter content of spacetime is described by a
distribution function $f$, but the vacuum case is included as the case
where $f$ is identically zero. Secondly, since the arguments used
there are local in space, the distinction between homogeneity and
local homogeneity is irrelevant.  All the results of the lemma for a
spacetime with Bianchi symmetry except the last (that $|H|$ tends to
zero) follow from Lemma 2.2 of \cite{AlanBianchi} and the discussion
immediately following it. The corresponding results for a spacetime
with Kantowski-Sachs symmetry follow from Lemma 3.1 of
\cite{AlanBianchi}, or can be derived by direct calculations in the
Schwarzschild space--time. It also follows that if $H$ remains bounded for
$\tau$ greater than some $\tau_0$ then $H<0$ for $\tau>\tau_0$. The
inequality $\partial_\tau H\ge{1\over 3}H^2$ ((2.35) in
\cite{AlanBianchi}) then implies that $H\to 0$ as $\tau\to\infty$. A
similar argument applies if $\tau$ is replaced by $-\tau$ and this
completes the proof  of the lemma.
\hfill $\Box$

Returning to the proof of Theorem \ref{TLH.1}, by Lemma
\ref{Alanlemma} the mean extrinsic curvature ${H}(s)$ of the slices
${\Sigma}_s = \{p \in {M} \mid {t}(p) = s\}$ tends in modulus either to
infinity or to zero as $s$ approaches $0$ or $-\infty$.  The
inextendability of $(M,g)$ in the class of globally hyperbolic
space--times in that time direction in which $H$ blows up follows now
from Hawking's focussing Lemma \cite[Prop.  4.4.3]{HE} (\cfeg
\cite[Prop.\ C.2.5]{SCC}).
It follows from \cite[Theorem 2.1]{RendallBianchiMatter}
 that the
space--time must be causally geodesically complete in that time
direction in which $H$ tends to zero (if any), from which
inextendability again follows.

To finish the proof of Theorem \ref{TLH.1}, we need to show
that all level sets of $t$ are Cauchy surfaces.  In the case of
compact $\Sigma$ this follows immediately from
\cite{BILY} (\cf also \cite{Galloway}).  In the general case the
arguments above show that this will follow if we show that the level
sets of $\hat{t}$ are Cauchy surfaces.  Consider then a surface
$\hat{\Sigma}_{t_0}$, by \cite[Prop.\ 2.4]{Chorbits}
$\stackrel{\circ}{{\cal D}}\!\!  (\hat{\Sigma}_{t_0};\hat{M})$ is the
maximal globally hyperbolic development of $(\hat{\Sigma}_{t_0},
\hat{\gamma}_{t_0}, \hat{K}_{t_0})$, where $(\hat{\gamma}_{t_0},
\hat{K}_{t_0})$ are the initial data induced on $\hat{\Sigma}_{t_0}$
from $\hat{g}$.  By \cite[Theorem 2.1.4]{SCC} $G$ acts on
$\stackrel{\circ}{{\cal D}}\!\! (\hat{\Sigma}_{t_0};\hat{M})$ by
isometries.  It is easily seen that $G$ acts transitively on all the
$\hat{\Sigma}_t$'s, so that we must have $\partial
\stackrel{\circ}{{\cal D}}\!\! (\hat{\Sigma}_{t_0};\hat{M})\cap
\hat{\Sigma}_t = \emptyset$ for all $t \in (-\infty, 0)$. Hence
$\partial\!\!  \stackrel{\circ}{{\cal D}}\!\!
(\hat{\Sigma}_{t_0};\hat{M}) = \emptyset$, which completes the proof.
\hfill $\Box$

Consider now $p \in \Sigma$, let $\hat{p} \in \pi^{-1}(p)$ and let
$\hat{{\cal O}}$ be a neighbourhood of $\hat{p}$ such that $\pi :
\hat{{\cal O}} \leftrightarrow {\cal O}$ is a diffeomorphism.  For any
Killing vector field $\hat{X} \in \Lie$ there exists a corresponding
Killing vector field $X = \pi_* X$ defined on ${\cal O}$.  For $g \in
G$ let $\hat{\phi}_g$ denote the action of $G$ on $\hat{M}$, for
$\epsilon > 0$ set
$$ {\cal O}(p, \epsilon) = \pi \left(\{\hat q =
\phi_{\exp(\hat{X})}(\hat{p}) : |\hat{X}| < \epsilon\}\right) .
$$ There exists $\epsilon(p) > 0$ such that for all $0 <
\epsilon \leq
\epsilon(p)$ the sets ${\cal O}(p, \epsilon)$ are connected,
simply connected open neighbourhoods of $p$.  We have the following:

\begin{Lemma} \label{LLH.1} Under the hypotheses of Theorem \ref{TLH.1},
suppose moreover that $\Sigma$ is compact.  Let $\Sigma_t$ be the
level sets of the function $t$ given by Theorem \ref{TLH.1}.  For every
$\epsilon > 0$ there exists $I(\epsilon)<
\infty$ such that for any set of points $\{p_i\}_{i = 1}^I \subset \Sigma_t$
there exist $i, j$ such that $$ {\cal O}(p_i, \epsilon) \cap {\cal
O}(p_j, \epsilon) \not= \emptyset\, .  $$
\end{Lemma}
{\bf Remark:} Let us emphasize that $I(\epsilon)$ given here is
{\em $t$--independent}.

\proof\ Let $\oB(p, s)$ denote the open geodesic ball of radius
$s$ centered at $p$, where the distance is measured with respect to
the metric $\gamma$ on $\Sigma \approx \Sigma_t$. (Here we use the
diffeomorphism $M\approx (-\infty,0)\times \Sigma$ to transport the
metric $\gamma$ from $\Sigma$ to $\Sigma_t$).  Choose $q_0 \in \Sigma$
and let $\epsilon_1(\epsilon)$ be such that $\oB(q_0,
\epsilon_1)$ is diffeomorphic to an open ball in $\Bbb{R}^3$, and that
$\oB(q_0,
\epsilon_1) \subset {\cal O}(q_0, \epsilon)$.  By $G$--invariance
of $\gamma$ it then follows that for all $p \in \Sigma$ we shall also
have $\oB(p,\epsilon_1) \subset {\cal O}(p, \epsilon)$.
{}From the covering $\{\,\oB(q, \epsilon_1/3)\}_{q \in \Sigma}$
of $\Sigma$ a finite covering $\{\,\oB(q_i,
\epsilon_1/3)\}^J_{i=1}$ can be chosen.

Let $I(\epsilon)$ be the largest integer larger than or equal to
$J/2$.  To see that $I(\epsilon)$ satisfies our assertions  consider any
set of points $\{p_i\}^I_{i=1}$.  Because the balls $\oB(q_i,
\epsilon_1/3)$ cover $\Sigma$, every point $p \in \Sigma$ is a
distance at least $\epsilon_1/3$ for some $q_{i(p)} $. Note, moreover,
that every ${\cal O}(p_i, \epsilon)$ contains a ball of radius
$\epsilon_1$, and therefore at least two points $q_i$.  It follows
that any collection of more than $\lceil J/2\rceil$ sets ${\cal O}(p_i,
\epsilon)$ has to have at least one intersection ${\cal O}(p_i,
\epsilon) \cap {\cal O}(p_j, \epsilon)$ non-empty.\hfill $\Box$

Let $G$ be the connected component of the identity of the group of
isometries of the space--times $(\hat M,\hat g)$ discussed in the
proof of Theorem \ref{TLH.1}, and let $\Lie$ be the Lie algebra of
$G$. We have the following:

\begin{Proposition}\label{PLH.3}
 Let $(\Sigma, \gamma, K)$ be locally
homogeneous vacuum initial data on a compact manifold $\Sigma$,
consider a space--time $(M, g)$ which is a (non--trivial) extension of
the maximal globally hyperbolic development of $(\Sigma, \gamma, K)$,
with a metric of differentiability class $C^{k,1}$, $k \geq 1$.  Then
for every $p \in \partial {\cal D}(\Sigma)$ there exists an open
neighbourhood ${\cal O}_p \subset M$ of $p$ such that
\begin{enumerate}
\item On ${\cal O}_p \cap \stackrel{\circ}{{\cal D}}\!\! (\Sigma)$
  there exists a Lie algebra $A_p$ of Killing vectors isomorphic to
  $\Lie$.

\item  The vector fields $X \in A_p$ extend $C^{k, 1}$ continuously to
$\partial {\cal D}(\Sigma)$.
\item  
The orbits of $X \in A_p$
through $p\in \partial {\cal
  D}(\Sigma)$ are three--dimensional.
\end{enumerate}
\end{Proposition}

{\bf Remark:} Point 3 above does not hold if the compactness condition
is relaxed. As an example, consider the unit spacelike hyperboloid in
Minkowski space--time with the corresponding induced Cauchy data. The
maximal globally hyperbolic development of the initial data is
isometrically diffeomorphic to the
interior of the light cone of the origin.  Taking the Minkowski
space--time as the extension, the Cauchy horizon is then the
light--cone of the origin, with the orbits of the symmetry group
through the origin being $0$--dimensional.  In this example, however,
there exists an open dense subset of the horizon on which the orbits
are three--dimensional. We believe that this is always true, even if the
compactness of $\Sigma$ is not assumed.

\proof\ Let ${\cal O}_p\subset M$ be any neighbourhood of $p\in
  \partial {\cal D}(\Sigma)$ such that \mbox{${\cal
      O}_p\cap\stackrel{\circ}{{\cal D}}\!\! (\Sigma)$} is connected
  and simply connected.  Choose some point \hbox{$q \in {\cal O}_p
    \cap \stackrel{\circ}{{\cal D}}\!\! (\Sigma)$}. As explained in
  the discussion preceeding Lemma \ref{LLH.1}, there exists a
  neighbourhood of $q$ and a Lie algebra of Killing vector fields
  isomorphic to $\Lie$ defined there.  A theorem of Hall \cite{Hall}
  (\cf also \cite{Nomizu,Singer}) shows that the Killing vectors can be
  extended to ${\cal O}_p\cap\stackrel{\circ}{{\cal D}}\!\! (\Sigma)$,
  and point 1 follows.  Point 2 follows as in the proof of Proposition
  \ref{P1}, Section \ref{BIX}.

  To establish point 3, let $K = \{ p \in \partial{\cal D}(\Sigma)\ |$
  the orbits of $x \in A_p$ through $p$ are {\it not}
  three--dimensional\}.  Consider a point $q \in K$, there exists a
  neighbourhood ${\cal U}_q \subset {\cal O}_q \cap\partial {\cal
  D}(\Sigma)$ of $q$ and $\epsilon > 0$ such that $\phi_{\exp(X)}(p)$
  is defined for all $p \in {\cal U}_q$ and $|X| < \epsilon$ by
  following the integral curve of $X$ through $p$ a
  parameter--distance one.  Clearly $\partial {\cal D}(\Sigma)
  \backslash K$ is open, 
we want to show that it is the whole of $\partial {\cal D}(\Sigma)$. In order
to do this consider $p \in \partial{\cal D}(\Sigma)$  such that the orbit
$$ {\cal O}(p, 5\epsilon) = \{ \phi_{\exp (X)}(p), \quad |X| < 5
\epsilon\} $$ is {\it not} three--dimensional.  Reducing
$\epsilon$ if necessary $ {\cal O}(p, 5\epsilon)$ is a smooth embedded
submanifold of $M$; further reducing $\epsilon$ if required we can find a
neighbourhood ${\cal O}$ of
$p$ such that $\phi_{\exp (X)}(q)$ is defined for all $q \in {\cal O}$
and for all $|X| \leq 4 \epsilon$.  We claim that there exists a
sequence $p_i \in {\cal O}$, $i \in \Bbb{N}$, such that
\begin{equation}\label{thething}
 {\cal O}(p_i, \epsilon) \cap {\cal O}(p_j , \epsilon) = \emptyset
\quad \hbox{if} \quad i
\not= j\, .
\end{equation}
 We shall proceed by induction.  Thus, let $p_1$ be any point in
${\cal O} \setminus {\cal O}(p, 5
\epsilon)$.  Suppose that a sequence
$p_i  \in {\cal O} \backslash {\cal O} (p, 5
\epsilon)$, $ i = 1, \ldots I$, satisfying \eq{thething} has been constructed
for some $I \geq 1$.  Set
$$
\Omega = \cup^I_{i = 1} \overline{{\cal O}(p_i , 2 \epsilon)} .
$$ We have $\Omega \cap \overline{{\cal O}(p, 2 \epsilon)} =
\emptyset$, which implies that there exists a neighbourhood ${\cal
O}_I{\subset} {\cal O}$ of $p$ such that $\Omega \cap {\cal O}_I =
\emptyset$.  As ${\cal O}(p_1 , 5 \epsilon)$ is a smooth embedded submanifold
of dimension less than three,
there exists a point
$p_{I+1} \in {\cal O}_I \backslash {\cal O}(p, 5\epsilon)$.  As
$p_{I+1} \not\in \Omega$,  one has for all $i = 1,\ldots I$
$$ {\cal O}(p_i, \epsilon) \cap {\cal O}(p_{I+1}, \epsilon) =
\emptyset \,,  $$ so that the induction step is completed.  [Clearly the
sequence $\{p_i\}$ can be so chosen so that $p_i \rightarrow p$, this will
however not be relevant in what follows.]

Let $T$ be any timelike vector field defined on ${\cal U}_q$, we
define $q_i(s)$ as the intersection of the orbit of $T$ through $q_i$
with the level sets $\Sigma_s$ of the function $t \in
C^\infty(\stackrel{\circ}{{\cal D}}\!\!  (\Sigma))$ given by Theorem
\ref{TLH.1}.  Here $t$ is chosen so that $t(p) \to 0$ as $p \to
\partial {\cal D}(\Sigma)$.  Let $I(\epsilon)$ be given by Lemma
\ref{LLH.1}, choose any set $\{q_i\}^{I(\epsilon)}_{i=1}$. Now the
sets $\{\phi_{\exp(X)}(q_i)\ |\  |X| < \epsilon\}$ are mutually disjoint,
and continuous dependence of solutions of ODE's upon initial values
shows that for $|s|$ small enough the sets
$\{\phi_{\exp(X)}(q(s))\,|\, |X| < \epsilon\}$ will be mutually
disjoint, which contradicts Lemma \ref{LLH.1} and proves the result.
\hfill $\Box$

{\bf Proof of Theorem \ref{T3}:} Suppose now that a spatially locally
homogeneous space--time with a compact partial Cauchy surface is of
Bianchi class A.  Then the obvious analogues of Lemma \ref{1} and
Lemma \ref{2} hold.  There exists a null frame for which the only
non--vanishing NP coefficients on the horizon are $\rho$ and $\gamma$.
Moreover $\RE(\gamma\bar\rho) = 0$ and $\RE\gamma > 0$.  The region
$S_1$, where $(\IM \gamma)(\IM\rho) < 0$ was discussed in the previous
section, {\em cf.\/} eq.\ (\ref{defS1}).  Consider now the following
subsets of $\C^{\myspace 2}$:
\begin{eqnarray*}
& S_2 = \{(\rho, \gamma) \in \C^{\myspace 2} : \RE (\gamma\bar\rho) = 0 ,\
\RE\gamma > 0 , \ (\IM\gamma)(\IM\rho) > 0\}\,,&\\
& S_3 = \{(\rho, \gamma) \in \C^{\myspace 2} : \RE
(\gamma\bar\rho) = 0 ,\  \RE\gamma > 0 , \ \IM\gamma = 0,\  \IM\rho
\not= 0\}\,,&\\
& S_4 = \{(\rho, \gamma) \in \C^{\myspace 2} : \RE (\gamma\bar\rho) =
0 , \ \RE\gamma > 0
,\  \IM\gamma \ne 0, \ \rho = 0\}\,, \\
& S_5 = \{(\rho, \gamma) \in \C^{\myspace 2} : \RE (\gamma\bar\rho) =
0 , \ \RE\gamma > 0
,\  \IM\gamma = \IM\rho = 0\}\,.
\end{eqnarray*}
Just as a point of $S_1$, can only arise from a space--time of type
IX, points of $S_2$, $S_3$, $S_4$ and $S_5$ can only arise from
space--times of types VIII, II, VII$_0$ and I respectively.  This can
be seen by applying the criteria$^{\ref{misprintsfootnote}}$ given by
Siklos \cite[Section 3]{SiklosCMP} for the different Bianchi types in
terms of the NP coefficients on the horizon. This identification of
the possible Lie algebras is already enough to give an interesting
conclusion concerning the Bianchi types of class A mentioned in
Proposition
\ref{P0}: there are
no Cauchy horizons in spatially locally homogeneous vacuum
space--times of Bianchi types VI$_0$ admitting a compact partial
Cauchy surface. What has been said above shows that this statement
remains true without the compactness condition as long as we assume
that the Cauchy horizon is locally homogeneous.  The vacuum
space--times of Bianchi type I are the (possibly twisted) Kasner
solutions.  It is known that most of these do not possess a Cauchy
horizon.  The only ones which do are those which are
flat\footnote{Those space--times have been overlooked in the list
given in \cite[Theorem 2]{SiklosCMP}}. The spin
coefficients in the set $S_4$ arise from these same space--times; the
Bianchi VII$_0$ frame is obtained from the Bianchi I frame by a
suitable rotation of $x$ and $y$.  More precisely, any { locally
homogeneous} metric of Bianchi I symmetry type can be represented as a
{ locally homogeneous} metric of Bianchi VII$_0$ symmetry type by
passing from a local basis of Killing vectors $\partial/\partial x$,
$\partial /\partial y$ and $\partial /\partial z$ to the locally
defined basis $\partial/\partial x$, $\partial /\partial y$ and
$y\partial/\partial x - x\partial /\partial y + \partial /\partial
z$. To complete the picture of which spatially compact class A
solutions admit Cauchy horizons it remains to study the type II and
type VIII solutions.  Space--times which realize the remaining values
of the NP coefficients can be obtained by making certain replacements
in the Taub--NUT metric. Suppose first that the vectors $X_1, X_2$ and
$X_3$ have the commutation relations $[X_1, X_2] = - X_3$, $[X_2, X_3]
= X_1$, $[X_3, X_1] = -X_2$ and replace $U$ by $-U$. The NP
coefficients of the resulting space--time differ on the horizon from
the Taub--NUT case only in that the signs of $\IM\gamma$ and $\RE\rho$
are reversed.  They realize the points of $S_2$.  To obtain the NP
coefficients in $S_3$ do a similar thing with the commutation
relations $[X_1, X_2] = 0$, $[X_2, X_3] = X_1$, $[X_3, X_1] = 0$ and
$U$ replaced by $\tilde U = {2mt\over t^2+L^2}$.  This time the NP
coefficients differ from those in the Taub--NUT case by the facts that
$\IM\gamma = \RE \rho = 0$ and that $U$ is replaced by $\tilde U$.
These are the generalized NUT space--times of types VIII and II
studied by Siklos \cite{SiklosPL}. (They can be compactified, as will
be shown below.)

It was shown in \cite{FIK} that for all spatially compact space-times
 with Bianchi B symmetry type the algebra of Killing vectors tangent
 to the surfaces of homogeneity must be at least four dimensional, and
 Theorem \ref{T3} follows.  \hfill $\Box$

To prove Proposition \ref{P0} we shall need several auxiliairy results.
The following result, which will be used later, has some
interest of its own:

\begin{Proposition}
\label{compactnesslemma} Let $(M,g)$ be a vacuum spacetime which is
the maximal globally hyperbolic development of locally homogeneous
initial data 
on a compact Cauchy surface
$\Sigma$. If $(M,g)$ has an extension $(M',g')$ through a future
Cauchy horizon then it has a possibly different extension $(M'',g'')$
through a smooth compact future Cauchy horizon.
\end{Proposition}

{\bf Remark:} The extension we are constructing  here
needs not to be vacuum beyond the new Cauchy horizon, even if
$(M',g')$ were.

{\bf Proof:} As discussed previously all vacuum globally hyperbolic locally
homogeneous
space--times with Kantowski--Sachs symmetry type are known, and the result
is true for those space--times, it is
therefore sufficient to consider the Bianchi symmetry
type. Let $G$ be the unique simply connected Lie group
corresponding to the given Bianchi type. Then there exists a discrete
group $\Gamma$ of tranformations of $G$ such that the spacetime $M$
can be identified with $G/\Gamma\times (-\infty,0)$. Moreover the
pullback of $g$ to $G\times (-\infty,0)$ is invariant under the action
of $G$ by left translations. If we denote a point of $G\times
(-\infty,0)$ by $(x,t)$ then $t$ can be chosen such that a future
Cauchy horizon in any extension of $(M,g)$ corresponds to $t\to
0$. Let $\gamma$ be a null geodesic in $(M',g')$ which crosses the
future Cauchy horizon at a point $p$. The identification of $M$ with
$G/\Gamma\times (-\infty,0)$ can be chosen so that the points of
$\gamma$ all correspond to the $\Gamma$-orbit of the identity of
$G$. It was shown above that the local action of $G$ on $M$ can be
extended to a local action of $G$ on the union $\cal V$ of $M$ with an
open neighbourhood of $p$ in the Cauchy horizon. This action can be
used to extend the identification of $M$ with $G/\Gamma\times
(-\infty,0)$ to an identification of $\cal V$ with ${\cal
W}=G/\Gamma\times (-\infty,0)\cup{\cal U}\times\{0\}$, where $\cal U$
is an open neighbourhood of the orbit of the identity.  Let $\hat{\cal
W}$ be the inverse image of $\cal W$ under the projection of $G\times
(-\infty,0]$ onto $G/\Gamma\times (-\infty,0]$.  Using the action of
the group the pull back to $\hat{\cal W}$ of the restriction of $g'$
to $\cal W$ can be extended to a $G$-invariant Lorentz metric on
$G\times (-\infty,0]$. It is also $\Gamma$-invariant by continuity
since the metric on the region $t<0$ is $\Gamma$-invariant.  Taking
the quotient by the group $\Gamma$ gives a ``space--time with
boundary'' $G/\Gamma \times (-\infty,0]$, which is clearly smoothly
extendable to some (perhaps non--vacuum) Lorentzian space--time
$(M'',g'')$ which contains a compact Cauchy horizon.
\hfill $\Box$

We shall also need the following lemma:

\begin{Lemma}\label{notflat}
Under the hypotheses of Theorem \ref{T3} suppose that $(M,g)$ is of
Bianchi B symmetry type. Then $(M,g)$ is {\em not} flat. Moreover the
group of isometries of the universal cover of $\ds$ contains a
subgroup of Bianchi type III.
\end{Lemma}

{\bf Proof:} By Lemma \ref{compactnesslemma} without loss of
generality we may assume that $\pds$ is compact and smooth.  It then
follows from \cite[p. 297]{HE} that $\mu|_{\pds}=0$. Lemma
\ref{Alanlemma} and the discussion before Lemma
\ref{2} show that $\Re\gamma|_{\pds}\ne 0 $. Recall now that
 for any
$B\in\C$ the ``gauge--transformation'' of the tetrad on $\pds$,
$$ n^a\to n^a, \quad m^a\to m^a+\bar Bn^a,\quad l^a\to l^a+Bm^a+\bar
Bm^{a}+B\bar Bn^a, $$ preserves the null--orthonormality
conditions. With a little work one finds that under this transformation
we have $$
\si\to\si+\bar B(\ta+2\beta)+\bar B^2
(\mu+2\bar\ga)+\bar B^3\nu.
$$
As
$(\mu+2\bar\ga)|_{\pds}\neq0$  we can achieve
$\si|_{\partial\ds}=0$ by performing such a transformation.
Hence without loss of generality we may assume that
$\si|_{\pds}= 0$. The equations \thetag{5.3}-\thetag{5.8} of
\cite{SiklosCMP} lead then to the following two sets of values of
NP coefficients
at $\pds$  (in each case below all
the unlisted NP coefficients vanish at $\pds$):
\begin{enumerate}
	\item
	\label{afamily}$
\rh\in\C\setminus\{0\} $, $\ga\in \R\setminus\{0\}$,
$\al\in\C $ \mbox{ subject to }  $
|\al|^2=1/3(\bar\ga\rh+\ga{\bar\rh})$, $
\ta=\bar\al $;
	\item  \label{afamily2}$
\rh\in\C\setminus\{0\} $, $\ga\in \R\setminus\{0\}$,
$\al\in\C $ \mbox{ subject to }  $
|\al|^2=1/4(\bar\ga\rh+\ga{\bar\rh})$, $
\beta=-\bar\al $;
\end{enumerate}
 (the case $\rh=0$ leads to a Bianchi class A symmetry type). Note that the
hypotheses of Proposition \ref{PU2} are again satisfied. The corresponding
metrics are therefore {uniquely} determined by the values (modulo gauges) of
the NP coefficients at $\pds$. It is easily checked that the metrics one
obtains in this way are of Bianchi III symmetry type. From eqs.\
\eq{B2}--\eq{(m)} one finds that the corresponding metrics are {\em not}
flat at the Cauchy horizon. (The two  families
 of boundary values of NP coefficients listed above are gauge--equivalent
 in the sense that they correspond to the same space--time
 metrics\footnote{We are grateful to
S. Siklos for pointing this out.}. This is, however, irrelevant for our
purposes.) \hfill$\Box$

There has been recently some interest in the global structure of
maximal globally hyperbolic flat Lorentzian space--times. Let us point
out the following:

\begin{Proposition}\label{flatLorenz}
Let $(M,g)$ be a spatially compact, locally homogeneous, maximal
globally hyperbolic $n+1$ dimensional flat Lorentzian manifold,
$n=1,2,3$. If the surfaces of homogeneity are not quotients of an
$n$--dimensional  torus, then $(M,g)$ is inextendible.
\end{Proposition}

{\bf Proof:} Multiplying $(M,g)$ by $S^1$ when $n=3$ or by $S^2$ when $n=2$
and equipping the resulting space--time with the obvious flat product
metric we can without loss of generality assume that $n=4$.
First the
possibility of Kantowski-Sachs symmetry will be eliminated. In that
case the universal covering spacetime is spherically symmetric. Define
a function $r$ by the condition that its value at any point is such that
the area of the orbit through that point under the action of $SO(3)$ is
$4\pi r^2$. Define the mass function by $m=(r/2)(1+\nabla_\alpha r
\nabla^\alpha r)$. The function $r$ is necessarily constant on the
hypersurfaces of homogeneity and so its gradient is timelike. Hence
$m$ is positive. However a direct calculation of the curvature of a
spherically symmetric spacetime shows that it can only be flat if
$m=0$. Hence
$(M,g)$ is a flat four dimensional Lorentzian manifold with Bianchi
symmetry type. Suppose that
$(M,g)$  is extendible.
By Proposition
\ref{compactnesslemma} we can assume that the extension is across a
compact Cauchy horizon. By Lemma \ref{notflat} the symmetry type must
be Bianchi A, and the result follows from the list of extendible
Bianchi A vacuum space--times given in the proof of Theorem \ref{T3}.
$\hfill \Box$

Some similar results have been established\footnote{\label{Messnote}G.\
Mess, private communication; {\em cf.\/} also \cite{Mess}.}  by G. Mess by
rather more involved techniques.

{\bf Proof of Proposition \ref{P0}:}
 For the Kantowski--Sachs models solutions with
compact
Cauchy horizons can be obtained by identifying $(t,r,\vec \omega)\in
\R\times (0,2m)\times S^2$
with $(t+a,r,R\vec \omega)$, where $a\in\R$ and $R\in SO(3)$
\cite{KoikeTanimotoHosoya}, in a Schwarzschild manifold. For the Bianchi I
symmetry type the relevant metrics are the flat Kasner metrics, which
can be spatially compactified if desired.  For
the Bianchi VII$_0$ type these are again the flat Kasner metrics, as
described in the proof of Theorem \ref{T3}.
For the Bianchi II, III=VI$_{-1}$, IV, VI --- IX
symmetry types these are the space--times listed in Theorem 2 of
\cite{SiklosCMP}. (Another family of Bianchi III space--times
extendible
across a Cauchy horizon (both compact and non--compact, if desired)
is given by  the family of ``pseudo--Schwarzschild'' solutions ({\em
cf.\ e.g.\/} \cite[p. 73, ``A2 solutions'']{EhlersKundt} and
\cite{Foster}).
 The needed Bianchi V space--time which is extendible across a
(non--compact) Cauchy horizon can be taken to be the interior of the
light cone of the origin in Minkowski space--time. It follows from Proposition
\ref{flatLorenz} that spatially compact quotients of this space--time
are inextendible.

The NUT spacetimes of Bianchi types VIII and II can both be spatially
compactified, as will now be shown. Let $(N,h)$ be a Riemannian
manifold, $P$ a principal $U(1)$ bundle over $N$ and $A$ a connection
on $P$. There exists a unique Riemannian metric on each fibre
invariant under the action of $U(1)$ such that the length of the fibre
is $2\pi$. Call this metric $k$. Given the above data and a positive
function $a$ on $M$, a Riemannian metric $\gamma$ on $P$ can be
defined as follows. If $X$ and $Y$ are horizontal let
$\gamma(X,Y)=h(\pi_*X,\pi_*Y)$, where $\pi$ is the projection of $P$
onto $N$. If $X$ is horizontal and $Y$ vertical let
$\gamma(X,Y)=0$. If both $X$ and $Y$ are vertical let $\gamma(X,Y)=a^2
k(X,Y)$. The action of $U(1)$ on $P$ coming from the principal bundle
structure is an action by isometries of $\gamma$. As an example of
this construction, consider the case where $(N,h)$ is 2-dimensional,
$P$ is the unit tangent bundle, $A$ is the Levi-Civita connection and
$a=1$.  Each isometry of $h$ induces an isometry of $\gamma$ in an
obvious way in this example. Moreover, these isometries are all
distinct from the $U(1)$ isometry group which all these metrics
possess and so the isometry group of $\gamma$ is at least one
dimension greater than that of $h$. Applying this construction to the
hyperbolic plane gives a metric which by the above has at least a
four-dimensional isometry group. Moreover the isometries of $\gamma$
which are induced by orientation-preserving isometries of the base act
transitively on $P$. Since the group of
orientation-preserving isometries of the hyperbolic plane has a Lie
algebra of Bianchi type VIII the metric $\gamma$ is homogeneous with
Bianchi VIII symmetry. Now we could have done the same construction
starting from a compact quotient of the hyperbolic plane and in that
case $\gamma$ is locally homogeneous with Bianchi VIII symmetry and is
defined on a compact manifold which is the total space of a circle
bundle over a compact surface. It also has an additional global
Killing vector given by the $U(1)$ action. In terms of appropriate
one-forms on $P$ invariant under the group the metric takes the form
$\sigma_1^2+\sigma_2^2+\sigma_3^2$ and $\sigma_2$, $\sigma_3$ vanish
on vectors tangent to the fibres. A smooth family of metrics may be
generated from this single one by multiplying $h$ by constant
conformal factors and taking general constant values of $a$ in the
above construction. This proves that the type VIII NUT metrics are
compactifiable in a variety of ways.

Consider next the type II metrics. They can also be compactified with
the aid of the above construction but unfortunately not using the unit
tangent bundle. (The unit tangent bundle of a torus is trivial and so
the above construction applied to a flat metric on a torus would give
Bianchi type I metrics.) Instead it is necessary to construct the
appropriate circle bundle by some other method. The type II NUT metric
is, in the coordinate form given by Siklos \cite{SiklosPL}, $$-\tilde
U^{-1}dt^2+(2l)^2\tilde U(d\psi+\theta d\phi)^2+
(t^2+l^2)(d\theta^2+d\phi^2)\,,$$ which is of the form taken by the
metrics arising from the above construction.  What needs to be checked
in order to verify that this metric really does arise from a circle
bundle over the torus is that the locally defined 1-form
$d\psi+\theta d\phi$ is the coordinate form of a globally defined
$U(1)$ connection $A$. This can be arranged, as follows from
Chern-Weil theory. (For information on the relevant part of this
theory see e.g. \cite[pp.114-121]{Woodhouse}.
 That reference talks
about Hermitian line bundles rather than circle bundles but it is easy
to see that the two things are equivalent.)  In order that the
one-form come from a globally defined connection it suffices that its
exterior derivative should define a global smooth 2-form and that this
2-form should satisfy an integrality condition. In the present case
the 2-form is $d\theta\wedge d\phi$, which is globally defined and
smooth on the torus obtained by identifying $\theta$ and $\phi$
periodically and the integrality condition can be arranged by an
appropriate choice of the coordinate volume of this torus. The time
variation of the metric can once again be accomodated by constant
conformal rescalings of the metric on the torus and the use of a
time-dependent $a$.
\hfill $\Box$

\section{Some uniqueness results for Taub--NUT space--times} \label{BIXV}

In this section we shall prove some results concerning the question of
uniqueness of extensions of the maximal globally hyperbolic region of
the Taub--NUT space--time. Let us start by
a construction which shows that uniqueness of solutions of the Einstein
equations fails whenever a Cauchy horizon occurs. This is true
 even if analyticity conditions on the metric and on the
space--time are imposed.
 Let then $(M,g)$ be a vacuum
extension of a maximal globally hyperbolic space-time $(M_0,g_0)$,
$M_0\subset M$, with Cauchy surface $\Si\subset M_0$, $\Si$ being a
partial Cauchy surface in $M$. Let ${\cal B}$ be any embedded two-sided
three-dimensional submanifold of
$M\setminus\overline{\cD(\Si;M)}$.  We shall moreover suppose that
$M\setminus\overline{{\cal B}}$ is connected.
Let $(M_a,g_a)$, $a=1,2$, be
two copies of $M\setminus\overline{ {\cal B}}$ with the metric induced from
$g$.
As ${\cal B}$ is two-sided, there exists an open neighbourhood
${\cal O}$  of
${\cal B}$ such that ${\cal B}$ separates ${\cal O}$ into two disjoint open
sets ${\cal O}_a$, $a=1,2$, with $\overline{{\cal O}_1}\cap \overline{{\cal
O}_2}=\overline{ {\cal B}}$,
${\cal O}_1\cap {\cal O}_2=\emptyset$. Let $\ps_a$ denote the natural
embedding of
${\cal O}_a$ into $M_a$. Let $M_3$ be the disjoint union of $M_1,M_2$ and
${\cal O}$, with the following identifications: a point
$p\in{\cal O}_a\subset{\cal O}$ is identified with $\ps_a(p)\in M_a$. It is
easily seen that $M_3$ so defined is Hausdorff.

We can equip $M_3$ with the obvious manifold structure and an obvious
metric $g_3$ coming from $(M_1,g_1)$, $(M,g_2)$ and
$({\cal O},g|_{\cal O})$. Note that if $(M,g)$ were real analytic, then
$(M_3,g_3)$
can be equipped with the obvious real analytic structure, with
$g_3$ --- real analytic with respect to this structure. Let finally
$(M_4,g_4)$ be any
maximal\,\footnote{\label{fnmaximal} \cfeg
    \cite[Appendix C]{SCC} for a proof of existence of space--times
    maximal with respect to some property. It should be pointed out
that there is an error in that proof, as the relation $\prec$ defined
there is not a partial order. This is however easily corrected by
adding the requirement that the isometry $\Phi$ considered there
restricted to some fixed three--dimensional hypersurface be the
identity.} vacuum (real analytic, if appropriate) extension of $(M_3,g_3)$.
Then $(M_4,g_4)$ is a maximal vacuum (perhaps real analytic) extension of
$(M_0,g_0)$ which clearly is not isometric to $(M,g)$.

The above allows one to construct many non-isometric maximal vacuum
(perhaps analytic) extensions of a given extendable maximal globally
hyperbolic space-time. Those extensions all have the property that
the resulting metric ``locally looks the same in all extensions'', the
lack of isometries coming from cut--and--paste games inflicted upon the
original extension. Intuitively speaking, two sufficiently nearby
observers living in
two such extensions which originate from the globally byperbolic
region will ``see the same metric''. For this reason it seems of
interest to present an
alternative construction in which this will not be the case. The
construction that follows will, however, not preserve analyticity of
the metric if the original metric is analytic.

Consider thus $(M_0,g_0)$, $(M,g)$, $\Sigma$, as before, and let
${\cal N}_a$, $a=1,2$, be two smooth, embedded, null hypersurfaces in
$M\setminus\cD(\Sigma;M)$ intersecting transversally along a
smooth two-dimensional submanifold $S$:
$\overline{{\cal N}_1}\cap\overline{{\cal N}_2}=S$, ${\cal N}_1\cap{\cal
N}_2=\emptyset$. Choosing
the ${\cal N}_a$'s ``small enough'' we ensure that all the conditions
in \cite{RendallCIVP} needed for the well--posedness of the
characteristic initial value problem in this setting are met. Let
$\phi_a$, $a=1,2$, denote the free characteristic initial  data
induced from $g$ on ${\cal N}_a$ as described in \cite{RendallCIVP},
let $\ps$ denote the corresponding data on $S$, {\em cf.\/} \cite{RendallCIVP}
for details. ``Making the ${\cal N}_a$'s smaller'' if necessary
there exists on open embedded submanifold ${\cal O}$ of
$M\setminus{\cal D}({\Sigma;M)}$ such that the metric
$g|_{\cal O}$ is
uniquely determined by $(\phi_a,\ps)$. Note that this is true
regardless of any global causality violations in $(M,g)$, as any Lorentzian
metric is causally well behaved on ``sufficiently small'' regions.

Consider now a set of free charateristic initial data
$(\tilde\phi_1,\phi_2,\ps)$, with $\tilde\phi_1\neq\phi_1$ but
$\tilde\phi_1-\phi_1$ vanishing to infinite order on $\overline{{\cal N}_1}$
near
$S$. Passing to subsets of the ${\cal N}_a$'s and ${\cal O}$ if necessary
there exists a solution $\tilde g$ of the vacuum characteristic
initial value problem in ${\cal O}$ with initial data
$(\tilde\phi_1,\phi_2,\psi)$ such that $g-\tilde g$ vanishes to
infinite order on $\overline{{\cal O}}$ near ${\cal N}_2$. Let now
$(M_1,g_1)$ be
obtained by glueing $(M\setminus{\cal O},g|_{M\setminus{\cal O}})$ with
$({\cal O},\tilde g)$ along ${\cal N}_2$ in the obvious way. The desired
extension of $(M_0,g_0)$ is obtained by taking any maximal vacuum
extension of $(M_1,g_1)$.

The above constructions show that uniqueness of solutions of the
Cauchy problem for the vacuum Einstein equations is lost whenever Cauchy
horizons occur, and the best one can hope for is at most some form of
uniqueness-up-to-boundary of such extensions\footnote{That is, there
can be no uniqueness beyond Cauchy horizons unless some further
conditions on the extensions are imposed. The problem is, that we do
not know any reasonable  conditions which would ensure either
existence or at least uniqueness of extensions beyond Cauchy
horizons.}. We shall show that such
a result is true for Taub--NUT space-times. Before doing that some
information about the isometry group of Taub--NUT space-times will be
needed. Let us define the Taub space-time as
$M=(t_-,t_+)\times L(p,1)$, where $L(p,1)$ is a lens space, with
the metric
\begin{equation}
\label{Eulermetric}
ds=U^{-1}dt^2-(2L)^2U(d\mu+\cos\th d\varphi)^2
- (t^2+L^2)(d\th^2+\sin^2\th d\varphi^2) \,.
\end{equation}
Here $\mu,\th,\varphi$ are Euler coordinates on $S^3$ (with some
suitable supplementary identifications for the $L(p,1)$'s with
$p\neq 1$). $U$ is given by eq.\ \eq{UBianchi}, and the  $t_{\pm}$'s
are zeros of
$U$. Note that if $m=0$, then the hypersurface $\Si_0=\{t=0\}$ is
totally geodesic; let us in that case denote by $T$ the isometry
which is a time-reflection across $\Si_0$. By considerations
involving uniqueness of maximal surfaces similar to those in
\cite{ChImaxTaubNUT} it is easily seen that for
$m\neq 0$ no time--orientation--reversing isometries of $g$ exist.

Recall, finally, that the map from $\R^4$ to itself defined by
$(x,y,z,w)$ $\to$ $(-x,-y,z,w)$ induces an analytic map from $L(p,1)$ into
itself. In the Euler coordinates this map takes the form $(\mu,\th,\varphi)
\to(-\mu,\th,-\varphi)$. This leads obviously to an isometry of $M$ into
itself, which we shall denote by $S$.

Let $G$ be the group of all isometries of $(M,g)$, let $G_0$ be the
connected component of the identity of $G$. It is well known that
$G_0$ acts transitively on each surface $\Si_\ta=\{t=\ta\}$. Moreover
the isotropy group of each $p\in\Si_\ta$ consists of rotations in the
planes perpendicular to the planes spanned by
$\frac{\partial}{\partial \mu}$ and $\frac{\partial}{\partial t}$, in
the coordinate system above. For $a,b=\pm$ let $(M^{ab},g^{ab})$
denote the Taub--NUT space-times described in \cite{ChImaxTaubNUT}. We have the
following:

\begin{Proposition}
\label{BIXG}
\begin{enumerate}
\item
 Let $G$ be the group of all isometries of a Taub space-time
$(M,g)$. If $m=0$, then $G$ has precisely four connected components,
$G_0$, $TG_0$, $SG_0$ and $STG_0$. If $m\neq 0$, then $G$ has
precisely two connected components, $G_0$ and $SG_0$.

\item Let $G$ be the group of all isometries of a standard Taub--NUT
space-time   $(M^{ab},g^{ab})$:

If $m=0$ and $ab=+$, then $G$
has precisely two connected components, $G_0$ and $STG_0$.

If $m=0$ and $ab=-$, then $G$
has precisely two connected components, $G_0$ and $TG_0$.

If $m\ne 0$, then $G$
has precisely one connected component, $G_0$, regardless of the value
of $ab$.
\end{enumerate}
\end{Proposition}

{\bf Proof:}
 1. Let $\phi$ be an isometry of $(M,g)$. By composing $\phi$ with $T$ if
necessary we may assume that $\phi$ preserves time-orientation. By
Lemma 3.4 of \cite{ChImaxTaubNUT} $\phi$ leaves the distribution $P$ of planes
spanned by $\frac{\partial}{\partial t}$ and
$\frac{\partial}{\partial \mu}$ invariant. By composing $\phi$ with
$S$ if necessary we may assume that $\phi$ preserves the orientation
of those planes.

Consider any $p\in\Si_t$, there exists $g\in G_0$ such that
$(g\phi)(p)=p$. Because $g\phi$ is an isometry that preserves the
distribution $P$, it must also preserve the distribution $P^{\perp}$ of
planes perpendicular to the planes $P$. Replacing $g$ by $hg$, where
$h\in G_0$ is an element of the isotropy group of $p$, we can achieve
$g\phi|_{P_p^{\perp}}=id|_{P_p^{\perp}}$. Now $g\phi$ preserves the
hypersurfaces of homogeneity $\Si_t$, hence
$(g\phi)^*\frac{\partial}{\partial t}\sim\frac{\partial}{\partial t}$;
in fact
$(g\phi)^*\frac{\partial}{\partial t}=\frac{\partial}{\partial t}$.
As $g\phi$ is a Lorentz transformation of $P_p$ into itself which
preserves orientation, we must have $g\phi|_{P_p}=id_{P_p}$. We have
thus shown that $g\phi|_{T_pM}=id_{T_pM}$, which together with
$(g\phi)(p)=p$ and standard results on isometries shows that
$g\phi=id$.

2. Let $\phi$ be an isometry of $(M^{ab},g^{ab})$ into itself. By
Proposition 2.1 of \cite{ChImaxTaubNUT} $\phi|_M$ is an isometry of
the Taub space $(M,g)\subset (M^{ab},g^{ab})$ into itself. The
extendability of isometries of the Taub region in $STG_0$ to
isometries of $(M^{ab},g^{ab})$ with $ab=+$ follows now from
construction and from eq.\ (8c) in \cite{ChImaxTaubNUT}, similarly for
isometries in $TG_0$ in the $ab=-$ case. The nonextendability of the
appropriate isometries in the remaining cases follows by
considerations similar to the proof of Theorem 3.1 of
\cite{ChImaxTaubNUT}.
\hfill $\Box$

{\bf Proof of Theorem \ref{T4}:} By proposition \ref{PLH.3} for any
$p\in\partial\ds $ the orbits of the extensions of the Killing vector
fields from $\ods$ to $\partial\ds$ are three-dimensional. It follows
that the arguments of the proof of Lemma \ref{3} apply for any
$p\in\partial\ds $, so that for any $p\in\partial\ds $ there exists
a neighbourhood ${\cal W}_p$ in $\overline{\ds }$ of $p$ and an
equivariant one-to-one isometry $i_p$ of ${\cal W}_p$ into a subset of
a standard Taub--NUT space time $(M^{ab},g^{ab})$. Let us show that
the $i_p$'s can be patched together to a single isometry on every
connected component of the Cauchy horizon, say $i^+$ on ${\cal H}^+$,
and $i^-$ on ${\cal H}^-$. Let $p$ be a point on the future Cauchy
horizon. Consider the set of pairs $({\cal W},\phi)$ where ${\cal W}$
is an open neighbourhood of $p$ in $\overline{\ds}\cap I^+(\Si)$ and
$\phi$ is an isometry of ${\cal W}$ onto an open subset of a Taub--NUT
space time. Now fix one of these pairs, say $({\cal {\cal
W}}_0,\phi_0)$, and restrict consideration to those pairs $({\cal
W},\phi)$ which agree with $({\cal W}_0,\phi_0)$ on a neighbourhood of
$p$.  Any two isometries of this type agree on the intersection of
their domains and so there exists a maximal element $({\cal
W}_{\max{}},\phi_{\max{}})$ in this class.  ${\cal W}_{\max}$ is by
definition open. Let $q$ be a point of its boundary. There exists a
neighbourhood ${\cal Z}$ of $q$ and an isometry  $\ps$ of ${\cal Z}$
onto a subset of Taub--NUT space time by Lemma \ref{3}. Let $r$ be a point
of the horizon in ${\cal W}_{\max}\cap {{\cal Z}}$. By composing with
an isometry of Taub--NUT spacetime it can be arranged that
$\ps(r)=\phi_{\max{}}(r)$. The construction of the local isometry in Lemma
\ref{3} also includes a null frame and the vectors $l$ and $n$ in this
frame are uniquely defined.  Moreover $m$ is defined up to
multiplication by $e^{i\th}$, $\th$ constant. Let $(l',n',m')$ and
$(l'',n'',m'')$ be the images of $(l,n,m)$ under $\phi_{\max{}}$ and $\ps$
respectively. The spin coefficients of $(l',n',m')$ have by
construction a special form and those of $(l'',n'',m'')$ are equal to
those of $(l',n',m')$. It follows that $l''=l'$, $n''=n'$ and
$m''=e^{i\th}m'$ for a constant $\th$. By composing $\ps$ with an
isometry of Taub--NUT space time which fixes $\ps(r)$ it is possible
to set $m''=m'$. After this has been done $\phi_{\max}$ and $\ps$ agree
on a neighbourhood of $r$ and hence together they define an isometry
on ${\cal W}_{\max{}}\cup{\cal Z}$, contradicting the definition of
${\cal W}_{\max{}}$. Hence ${\cal W}_{\max}$ has no boundary point and,
since the future Cauchy horizon is connected, ${\cal
W}_{\max}=\overline{D(\Si)}\cap I^+(\Si)$. By equivariance the
$i^\pm$'s can be extended to a one-to-one isometry, still denoted by
$i^\pm$, from $\cD (\Sigma,M)$ into $(M^{ab},g^{ab})$. If ${{\cal
H}}^+$ or ${{\cal H}}^-$ is empty we are done, if not, consider the
map $j=i^+\circ (i^-)^{-1}$ which is an isometry from the Taub region
of $(M^{ab},g^{ab})$ into itself. Without loss of generality we may
assume that $j$ preserves time orientation. If $j\in G_0$ then the
isometries $i^t$ and $i^-$ can be patched together to a one-to-one
isometry-up-to boundary of $\overline{\cD (\Sigma,M)}$ into
$(M^{++},g^{++})$. Similary if $j\in SG_0$, then $i^+$ and $i^-$ can
be patched together to a one-to-one isometry up-to-boundary of
$\overline{\cD(\Sigma,M)}$ into $(M^{+-},g^{+-})$, and the result
follows. \hfill $\Box$

Our next result proves uniqueness of the standard Taub--NUT
space--times, in the class of maximal space--times with an action of
$G = SU(2)$ or $SO(3)$ by isometries:

\begin{Theorem} \label{TBV.I}  Let $(M,g)$ be a vacuum Bianchi IX
space--time with a compact partial Cauchy surface $\Sigma$ and
  nontrivial Cauchy horizon, $\partial {\cal D}(\Sigma)\ne \emptyset$.
  Suppose, moreover, that $(M, g)$ is maximal\,$^{\mbox{\ref{fnmaximal}}}$
  in the class of vacuum
  space--times on which $G = SU(2)$ or $SO(3)$ acts by isometries.
  Then $(M, g)$ is a standard Taub--NUT space--time.
\end{Theorem}

\proof\ Let $\Sigma$ be a partial Cauchy surface in $M$, by
Theorem \ref{T4} there exists a one--to--one map $i$ which maps
$\overline{{\cal D}(\Sigma; M)}$ into $\overline{{\cal D}(i(\Sigma),
M')}$, and which is a smooth isometric diffeomorphism between
$\stackrel{\circ}{\cal D}(\Sigma; M)$ and
$\stackrel{\circ}{{\cal D}}\!\! (i(\Sigma), M')$.  Here $M'$ is the
corresponding standard Taub--NUT space--time.  By maximality and
homogeneity $i$ has to be onto. The result follows by a Siklos tetrad
construction argument, as in Section \ref{BIX}, and by simple
maximality arguments. \hfill $\Box$

For  other Bianchi types standard
extensions can be constructed for each extendible Bianchi model, as
should be clear for the explicit form of the metrics given in
\cite{SiklosCMP,SiklosPL} or as discussed in Section \ref{LH}. Then the
obvious
equivalents of Theorems \ref{T4} and \ref{TBV.I} hold, the details are
left to
the reader. Note that it follows from Lemma \ref{Alanlemma} that future
and past Cauchy horizons occur simultaneously only for the Bianchi IX
models, so that the equivalent of Proposition \ref{BIXG} is not needed in
the non-Bianchi-IX cases.

Let us finally mention that uniqueness of standard extensions of
Kantowski-Sachs models does not follow immediately from what was said
above, though we believe that the arguments given could be adapted to
obtain such results. We have, however, not made attempts to analyze this
problem.

\ifx\undefined\bysame
\newcommand{\bysame}{\leavevmode\hbox to3em{\hrulefill}\,}
\fi

\end{document}